\documentclass[pra,twocolumn,showpacs,amsmath,amssymb,superscriptaddress,longbibliography]{revtex4-1}
\usepackage{psfrag,graphicx}
\usepackage{amsfonts,amssymb,amsmath,mathtools}      
\usepackage{graphicx}
\usepackage{dcolumn}
\usepackage{bm}
\usepackage{float}
\usepackage{hyperref}
\usepackage{accents}
\usepackage{bbding}
\usepackage{twemojis}
\usepackage{color,soul}
\usepackage{epstopdf}
\usepackage{changepage}
\usepackage[capitalize,noabbrev]{cleveref}
\crefname{section}{Sec.}{Secs.}
\Crefname{section}{Sec.}{Secs.}

\crefname{subsection}{Sec.}{Secs.}
\Crefname{subsection}{Sec.}{Secs.}
\usepackage{orcidlink}

\usepackage[normalem]{ulem}

\usepackage{xr}

\newcommand{\erf}[1]{Eq.~(\ref{#1})}
\newcommand{\beq}{\begin{equation}}
\newcommand{\eeq}{\end{equation}}
\newcommand{\nn}{\nonumber}

\newcommand{\crf}[1]{Ref.~\cite{#1}}

\newcommand{\neswarrow}{\rotatebox[origin=c]{135}{$\updownarrow$}}
\newcommand{\bra}[1]{\langle{#1}|}
\newcommand{\ket}[1]{|{#1}\rangle}

\newcommand{\dbd}[1]{\frac{\partial}{\partial {#1}}}

\newcommand{\cu}[1]{\left\{ {#1} \right\}}
\newcommand{\ro}[1]{\left( {#1} \right)}
\newcommand{\an}[1]{\left\langle{#1}\right\rangle}
\newcommand{\Tr}{\text{Tr}}

\newcommand{\s}[1]{\hat{\sigma}_{#1}}

\newcommand{\dd}{\text{d}}

\newcommand{\eg}{{\em e.g.}}

\newcommand{\ea}{{\em et.~al.}}

\newcommand{\union}{\cup}
\newcommand{\past}[1]{\overleftarrow{#1}}%\overleftarrow{#1}}
\newcommand{\fut}[1]{\overrightarrow{#1}}
\newcommand{\both}[1]{\overleftrightarrow{#1}}

\newcommand{\inv}{^{-1}}

\newcommand{\nullset}{{\emptyset}}

\renewcommand{\Pr}{{\rm Pr}}
\newcommand{\Su}{{\rm Su}}

\definecolor{nblue}{rgb}{0.3,0.3,1.0}%229
\definecolor{ngreen}{rgb}{0,0.7,0}%272%161
\definecolor{nred}{rgb}{0.7,0.1,0.1}%711&900%1022
\definecolor{nyellow}{rgb}{0.9,0.7,0.2}%772
\definecolor{npurple}{rgb}{0.6,0.0,0.6}%606
\definecolor{nrose}{rgb}{0.5,0.3,0.3}
\definecolor{nbackground}{rgb}{1,1,1}
\definecolor{ngrey}{rgb}{0.5,0.5,0.5}
\definecolor{nbrown}{rgb}{0.5,0.4,0.0}
\definecolor{nblack}{rgb}{0,0,0}
\definecolor{ncyan}{rgb}{0.1,0.5,0.5}
\definecolor{npale}{rgb}{0.9,0.9,0.9}
\definecolor{sky}{rgb}{0,0.6,0.8}%159
\definecolor{nmauve}{rgb}{0.1,0.0,0.7}
\definecolor{norange}{rgb}{1.0,0.4,0}
\definecolor{napricot}{rgb}{1,0.6,0.5} %10,8,7
\definecolor{golden}{rgb}{0.75,0.6,0.15}
\definecolor{nbrick}{rgb}{0.65,0.25,0.15}

\definecolor{nmagenta}{rgb}{0.7,0.0,0.3}

\newcommand{\blk}{\color{nblack}}

\newcommand{\cfg}{\,\|\,}%{\,@\,}

\newcommand{\frf}[1]{Fig.~\ref{#1}}

\usepackage[autostyle, english = american]{csquotes}
\MakeOuterQuote{"}

\begin{document}

\title{Counterfactual quantum measurements}

\author{Ingita Banerjee}
\affiliation{Centre for Quantum Computation and Communication Technology 
(Australian Research Council), \\ Quantum and Advanced Technologies Research Institute, Griffith 
University, Yuggera Country, Brisbane, Queensland 4111, Australia}

\author{Kiarn T. Laverick\orcidlink{0000-0002-3688-1159}}
\affiliation{MajuLab, CNRS-UCA-SU-NUS-NTU International Joint Research Laboratory}
\affiliation{Centre for Quantum Technologies, National University of Singapore, 117543 Singapore, Singapore}
%\affiliation{Centre for Quantum Computation and Communication Technology 
%(Australian Research Council), \\ Quantum and Advanced Technologies Research Institute, Griffith 
%University, Yuggera Country, Brisbane, Queensland 4111, Australia}

\author{Howard M. Wiseman\orcidlink{0000-0001-6815-854X}}
\email{h.wiseman@griffith.edu.au}
\affiliation{Centre for Quantum Computation and Communication Technology 
(Australian Research Council), \\ Quantum and Advanced Technologies Research Institute, Griffith 
University, Yuggera Country, Brisbane, Queensland 4111, Australia}

\date{\today}

\begin{abstract}

Counterfactual reasoning plays a crucial role in exploring hypothetical scenarios, by comparing some consequent under conditions identical except as results from a differing antecedent. 
%They are claimed to represent the highest form of reasoning about causality. 
David Lewis' well-known analysis evaluates counterfactuals using a hierarchy of desiderata. These were, however, built upon a deterministic classical framework, and whether it could be generalized to indeterministic quantum theory has been an open question. In this paper, we propose a formalism for quantum counterfactuals in which antecedents are %\ing{different?} 
measurement settings. Unlike other approaches, it non-trivially answers questions like: "Given that a photon-detector, observing an atom's fluorescence, clicked at a certain time, what would a field-quadrature detector have measured, if it had been used instead?"

\end{abstract}

\maketitle

\section{Introduction}
Counterfactual reasoning has applications in diverse fields of study including policy making~\cite{tetlock1996counterfactual}, game theory~\cite{epstude2008functional}, weather and climate~\cite{otto2017attribution}, artificial intelligence~\cite{celar2023people}, philosophy~\cite{lewis1979counterfactual,butterfield1992david,pearl2018book} and causal discovery~\cite{pearl2009causal,Imbens_Rubin_2015,wang2025quantumexperimentjointexogeneity,repec:ris:nobelp:2021_002}. 
Counterfactual statements also played an important role in development of quantum theory, 
as the 1935 debate between Einstein-Podolsky-Rosen (EPR)~\cite{PhysRev.47.777} and Bohr~\cite{PhysRev.48.696} 
famously centred around a counterfactual: the result of a measurement that could have been performed instead of the one that actually was.  
Despite their role in debates pertaining to EPR and Bell-nonlocality ---  
in particular whether `counterfactual definiteness' (that counterfactual propositions have definite truth values) was an implicit assumption in the derivation of Bell's theorem~\cite{skyrms1982counterfactual,PhysRevD.3.1303,lambare2021note,Zukowski2014} --- 
 the meaning and status  
of counterfactuals in quantum theory has remained unclear \cite{stapp1997nonlocal,PhysRevA.59.126,mermin1998nonlocal,stapp1998meaning,AHARONOV2002130,vaidman2009counterfactuals,vaidman2007counterfactualsquantummechanics}. 
%and much in part because such discussions often largely because counterfactual reasoning often relied on counterfactual definiteness
%Subsequently, there have been many debates . \red  and therefore running into  problems of  hidden variable like assumptions. \blk 

In this paper, we  aim to put quantum counterfactuals on a firm footing.~We are not concerned with %concepts like 
`counterfactual definiteness', which makes sense only in the context of deterministic hidden variables~\cite{RevModPhys.38.447,PhysRev.108.1070}. 
%(as EPR desired to `complete' quantum theory). 
Rather, we tackle the problem within the framework of orthodox quantum theory (OQT), 
%with associated ideas about causal structures but 
%without hidden variables. That is, 
which means grappling with the indeterminacy, and violation of classical causality~\cite{PhysicsPhysiqueFizika.1.195,Wood_2015,wiseman2016causarum,Kochen1990,Bell:609719}, in OQT. This %makes it a challenge for formulating counterfactuals, 
is challenging  
because %all of 
the best-developed philosophical analyses of counterfactuals %has been done 
assume classical systems with deterministic causation. Nevertheless, building on these analyses, we %will show that some of these analyses can be generalized for OQT in a way 
propose a counterfactual quantum calculus  
that lets us answers even very complicated %counterfactual 
questions like  "Given that my photon-detector, observing an atom's fluorescence, clicked at a certain time,  what would I have seen had I been using a field-quadrature detector instead?"
%\ing{We could change the sentence above a bit by saying we propose our calculus..thus extending the concept of counterfactuals beyond phenomenological arguments to ...}

The structure of this paper is as follows. In \cref{sec2:CC} we begin by briefly presenting the most useful classical framework for our purposes, namely that of Lewis~\cite{lewis1979counterfactual}.
This leads to a discussion of how to extend it to the quantum domain, including evaluating counterfactual probabilities rather than counterfactual propositions in \cref{sec3:Lewis}. In \cref{sec:Simple Eg.}, we use the example of a Bell-CHSH scenario to introduce our key ideas for evaluating non-trivial counterfactual probabilities in OQT by 
%carefully specifying and justifying the fixture of certain 
keeping fixed certain elements across both the actual and the counterfactual worlds. We then formalize these ideas in \cref{sec5:Calculus}, into a general methodology for counterfactual settings in OQT. Lastly, in \cref{sec6:}, we demonstrate its effectiveness by applying it in a highly non-trivial situation involving continuous monitoring of a resonantly driven atom by two observers. Our formalism bridges the gap between the power of quantum theory and the intuitive counterfactual reasoning used in scientific explanation and prediction.

%and potentially to other inherently probabilistic theories too~\cite{}.

%However, the ability  to talk about counterfactual scenarios in quantum mechanics is not straightforward since such definiteness isn't allowed owing to the probabilistic nature of quantum measurements.

\section{Classical Counterfactuals \label{sec2:CC}} Of the many approaches to evaluating counterfactuals in the classical realm, one of the most influential is that of Lewis~\cite{lewis1979counterfactual}. His approach is based on similarity analysis of possible worlds, where~\cite{lewis1979counterfactual} 
\enquote{a counterfactual `If it were that A, then it would be that C' is (non-vacuously) true if and only if some (accessible) world where both A and C are true is more similar to our actual world, overall, than is any world where A is true but C is false.}  Here A is the antecedent and C is the consequent. Lewis claims that the world most similar to the actual one can be found by following a hierarchy of desiderata, of which only the top two concern us:
\begin{enumerate}
    \item Avoid big, widespread, diverse violations of law.
    \item Maximize the spatio-temporal region throughout which perfect match of particular fact prevails.
   % \item Avoid even small, localized simple violations of law.
    %\item No importance to secure approximate similarity in concerning matters.
\end{enumerate}

This approach, has, faced criticism~\cite{Percival1999-PERANO-2,Elga2001-ELGCDA,horwich1987asymmetries,suresh2023semantics}, particularly %regarding the vagueness of 
%the vagueness of the ``similarity'' relations~. In particular, 
that it is not always clear as to what should remain ``matched'' when proposing the antecedent. 
%\ing{to maintain perfect match in particular facts with the actual world without violating any laws in it}.\hmw{You don't think my wording is clear enough? It introduces an idea that will be necessary later.} 
In addition to this, Lewis's analysis faces two major hurdles when it comes to applying it to  OQT~\cite{butterfield1992david}. First, it assumes a classical ontology, where there is a matter of fact about the physical state of the world at any point in space-time. Second, it assumes deterministic laws of time-evolution for those facts. This is why, for Lewis, a counterfactual proposition is either true or false. In OQT, on the other hand, most physical quantities do not have determinate properties, and the theory is inherently probabilistic when it comes to those that do (measurement results). %\ing{Lewis's original extension of his analysis to indeterministic theories~\cite{Lewis1986-LEWCE} was based on the idea to show that even in a chancy world, the world where the consequent is true is closer to the actual world according to a Refined ranking that incorporates the idea of indeterminism, but this faced criticism~\cite{Percival1999-PERANO-2}. Even in the indeterministic regime he still classified the counterfactuals as true or false. Here we relax this conservative approach and start to talk about chances of the occurrence of the consequent ( which mirrors the probabilistic measurement results in QT) rather than chances of the world where the consequent is guaranteed to be true.} 

%The reason is that the analysis was originally formulated in a deterministic framework, assuming classical ontology in space-time, which implies that each variable at a definite location has a well-defined value, even if it is not measured. 
%\hmw{Re ``These challenges make it very difficult to straightforwardly apply the analysis to QT. In fact an attempt to do so might lead to real problems~\cite{butterfield1992david}.'', I guess we don't necessarily disagree with Butterfield. We don't just apply Lewis's approach. Would it be better to leave these sentences out and cite Butterfield earlier in the paragraph?} 

 \section{Extending Lewis's analysis for Orthodox Quantum Theory \label{sec3:Lewis}}We now present four steps (restrictions, a generalisation and an  interpretation) that takes us from Lewis's approach to a formalism for counterfactuals in OQT. 
i) We restrict our description to the classical elements of the theory: measurement settings, measurement outcomes, and uncontrolled classical variables (noise). ii) We generalize from counterfactual propositions to counterfactual probabilities. 
That is, given the antecedent, rather than asking 
%We consider counterfactual expectations or what we call suspectations, rather than propositions i.e., what we calculate is not 
whether the consequent is true, we ask what is the probability (\Pr) of the consequent. 
%\ing{Or just the probability of the consequent? Because this sounds a bit similar to finding the probability of a world where the counterfactual proposition is true which was Lewis original idea of extension a bit.}
%, or, if the consequent refers to a variable, say $c$, what is the expectation of that variable, ${\mathbb E}[c]$. 
To avoid the potential misunderstanding that 
%these probabilities and expectations 
this counterfactual probability be straightforwardly empirically observable, we will instead  use the similar-sounding neologism  {\em supposability} (\Su). 
%and {\em suspectations} (${\mathbb S}[c]$). 

We note that steps (i) 
%\ing{Not sure if step 1 is similar to Ardra's work.}\hmw{Why not? In terms of the variables that are updated or estimated, it is only classical ones. The quantum causal model just specifies how these are connected, really. Do they not say this clearly? All of the variables you discuss in your summary of their work corresponds to classical variables, no?} 
and (ii) above (but not the new terminology) were also adopted in another recent work on 
%``a semantics for 
quantum counterfactuals~\cite{suresh2023semantics}. Our remaining steps are quite different, leading to radically different results; see \cref{Sec:Otherworks}.  iii) We address Lewis's first desideratum by restricting the antecedent to be measurement settings. In the context of OQT, settings are free choices external to the quantum systems under study, so there is no violation of law to consider a different choice. This type of counterfactual is also one of the most common and interesting types of speculative thinking by humans. 
iv) We interpret Lewis's second desideratum as requiring us to keep fixed (in actual and counterfactual worlds) 
%The choice of the antecedent and the identification of the fixed events based on the context is a primary step for counterfactual analysis and so we define our antecedent and the fixed events according to the above resolution which we will keep consistent for the rest of our discussion, (i) \textbf{Antecedent: Choice of some measurement settings.}     Following from Lewis' first desiderata it is clear that there is \textit{no} violation of laws of QT in changing the choice of the measurement. (The most interesting type of antecedent.) (ii) \textbf{Fixed events: 
any classical variables uninfluenced by the counterfactual antecedent. This means other measurement settings and outcomes, even if unknown to the person posing the counterfactual (or, to be more formal, even if unspecified by the theory or as ``evidence''~\cite{pearl2018book} in the counterfactual). 

Step (iv) of course raises the question of what outcomes are influenced by a setting in OQT. One could give a general answer in the framework of quantum causal models~\cite{Costa_2016,PhysRevX.7.031021,chaves2015information,PhysRevA.88.052130}, which fits well with the ontology of OQT. Adopting the appropriate quantum DAG (directed acyclic graph) for a given quantum experiment, a classical variable would be uninfluenced by a counterfactual setting if it is not amongst its descendants in the DAG. In this Letter, we avoid that complication by considering experiments with a space-time arrangement such that the variables uninfluenced by a counterfactual setting are so by virtue of being outside its future light cone. %Finally, we note that Lewis's third and fourth desiderata  are trivially satisfied in our formalism, and so need not detain us.

\section{A simple example \label{sec:Simple Eg.}} 
\begin{figure}
vs     \includegraphics[width=0.50\textwidth]{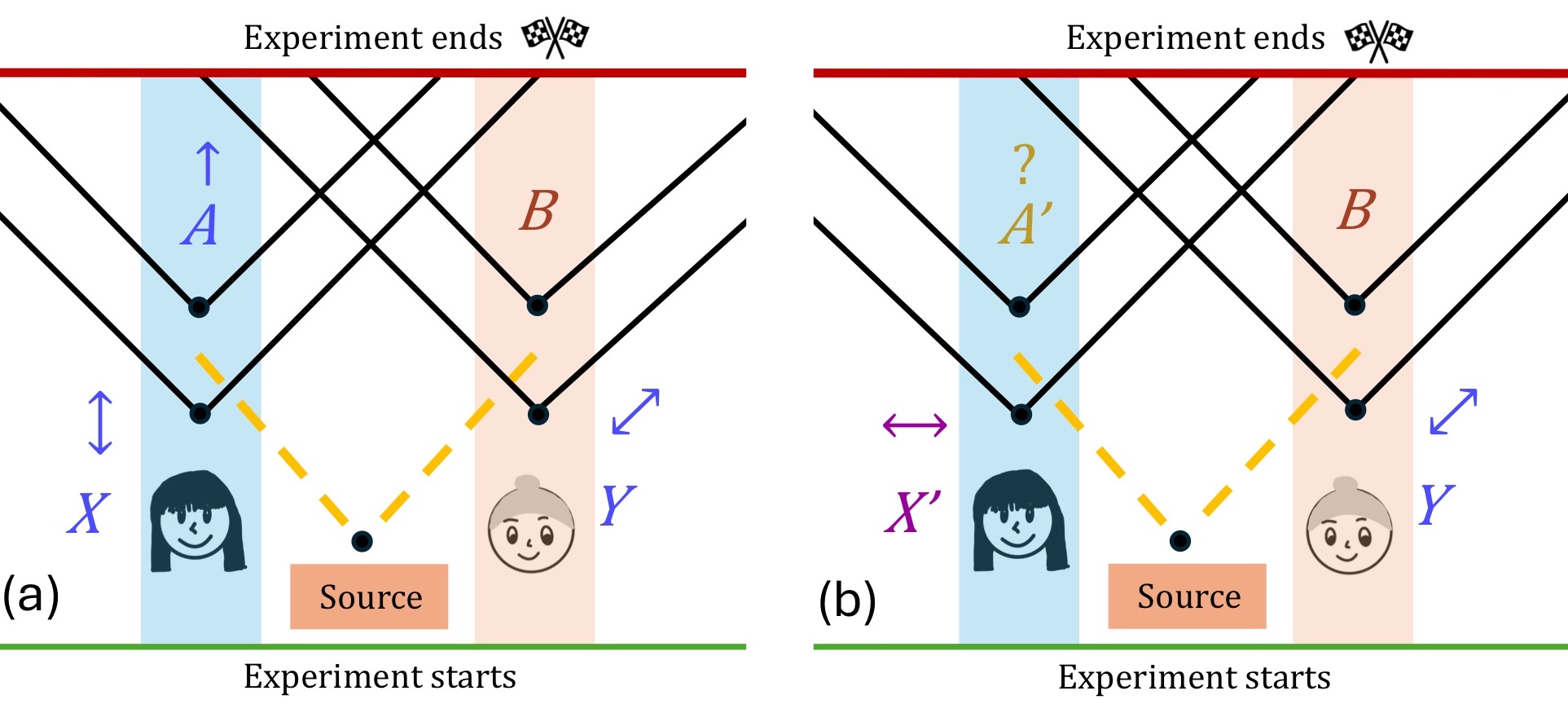}
    \caption{A counterfactual in the CHSH scenario. Alice and Bob make measurements on a shared singlet state. (a) Alice knows her outcome $A$ for the setting $X$, and also Bob's setting (blue, evidence). (b) Alice wonders about her outcome $A'$ (golden, consequent) if she had used a different setting $X'$ (purple, antecedent). In both cases Bob's outcomes $B$ (brown) is unknown to Alice but fixed. The black lines show a future light cone originating at an event in space-time.}
  \label{fig:sidebyside}
\end{figure}

Having explained the essential ideas of our formalism for quantum counterfactuals, we apply it to a simple example involving CHSH-type correlations, as illustrated in Fig.~\ref{fig:sidebyside}. 
We consider a standard Bell-CHSH experiment~\cite{PhysicsPhysiqueFizika.1.195,PhysRevLett.23.880}, where Alice and Bob are in distant laboratories and they perform  measurements on a shared spin-singlet state. We thus have four binary classical variables: 
    $X$ and $A$ for Alice's measurement setting and outcome, and, similarly, $Y$ and $B$ for Bob. Note that $(X,A)$ is space-like separated from $(Y,B)$. 
    We will use double-headed arrows for settings (\eg, 
    $X=\,\updownarrow$ means Alice measures spin in the $z$-direction), and single-headed arrows for outcomes (\eg, $A=\,\uparrow$ means she got the `up' result). 
    OQT supplies the  joint probability distribution of the outcomes given the settings as %\vspace{-0.5mm}
\begin{equation} \label{quantum_probabilties}
    \wp(a,b|x,y)=\Pr(A=a,B=b|X=x,Y=y)\,.
\end{equation}

Next, we identify the role of the various elements just introduced. To aid intuition, we consider a counterfactual question that would naturally occur to Alice at the end of one actual run of the experiment (as defined in Fig.~\ref{fig:sidebyside}), but we note that the question could also be posed abstractly. Alice at this time has certain \textbf{evidence}:
%(a term from Pearl~\cite{pearl2000models}): 
her own measurement setting $X=\,\updownarrow$, her outcome  $A=\,\uparrow$, and Bob's setting  $Y=$\rotatebox[origin=c]{135}{$\updownarrow$}. 
%(Basis rotated by 45 degrees from the Z-axis).\\
Alice wonders, 
``what would I have seen, had I chosen the other setting used for this CHSH experiment?''. That is, consistent with our step (iii), the 
\textbf{counterfactual antecedent}  
%At the end of the experiment Alice wonders an alternative scenario where 
is $X'=\,\leftrightarrow$, and the {\bf counterfactual consequent} is $A'$. (Note that we always use primed variables for the counterfactual world.) Now we can apply our step (iv): all classical events not in the future light-cone of $X'$ are {\bf fixtures}. This includes $Y$ (part of the evidence and so known) and $B$ (not part of the evidence and so unknown). 

%\textbf{Fixed events:} She has to decide what are the events that are to be kept fixed in this scenario and following from our definition of the fixed events she concludes that since the outcome of Bob lie outside the future light cone (FLC) of the antecedent event (the event of her measurement setting) it cannot be influenced by change in Alice's measurement settings and therefore should be held fixed, 
%\begin{equation}
%    B \notin \text{FLC}(X')\,.
%\end{equation}

%\textbf{The counterfactual proposed by Alice:} 
Having categorized the elements, Alice should, according to step (ii) of our formalism, frame her question as follows: ``Given the evidence $X=\,\updownarrow$, $A=\,\uparrow$, and $Y=$\rotatebox[origin=c]{135}{$\updownarrow$}, what is the probability that, had I chosen $X'=\,\leftrightarrow$, I would have seen  $A'=\,\rightarrow$?'' We notate this probability, or rather supposability, as 
\beq 
\Su(A'=\rightarrow | X'=\leftrightarrow || A=\uparrow,X=\updownarrow,Y=\rotatebox[origin=c]{135}{$\updownarrow$}), \label{supposability1}
\eeq
Note the two different types of conditioning here. The usual single bar ($|$) conditions the supposability of the result $A'$ in the counterfactual world where Alice's setting is $X'$. The double bar ($||$) conditions it on the evidence, comprising events in the actual world. The only events shared by the actual and counterfactual worlds are the fixtures. Thus, to evaluate \erf{supposability1}, we first determine the probability of the unknown fixtures (here, $B$) in the actual world, given the evidence. 
%Because $Y$ is part of the evidence, the only unknown fixture to consider is $B$. 
Then we can evaluate the counterfactual-world probability with the fixtures added to the conditional, and average with respect to $B$. That is, simplifying the notation by suppressing the values of the variables except $b \in \{\rotatebox[origin=c]{135}{$\uparrow$},\rotatebox[origin=c]{135}{$\downarrow$}\}$, 
by \erf{supposability1} we mean 
 \begin{align}
% \Pr(A'| X' || A,X,Y) %\label{CFP}
 \sum_{b} \,&\, \Pr(A'|B=b,X',Y) \Pr(B=b|A,X,Y) \nonumber \\
 &=\frac{\sqrt{2}+1}{2\sqrt{2}}\times\frac{\sqrt{2}+1}{2\sqrt{2}}+\frac{\sqrt{2}-1}{2\sqrt{2}}\times\frac{\sqrt{2}-1}{2\sqrt{2}}=\frac{3}{4},
 \label{eq:final}
 \end{align}
The numbers appearing in the final line all arise from the usual quantum probabilities (\ref{quantum_probabilties}), which we will discuss shortly. This supposability of $\frac34$ is an example of a non-trivial (as promised) answer to a counterfactual question. It 
differs from other obvious answers Alice might give, such as: ``If the setting is different then the evidence from this run is irrelevant. I should simply recalculate $\Pr(A'=\rightarrow | X'=\leftrightarrow)$,'' which equals $\frac12$; or ``If I got the result $A=\uparrow$ in the actual world, the state I received must have been $\ket{\uparrow}$. Thus, if I had measured $X'=\leftrightarrow$, I would have got the result $A'=\rightarrow$ with probability $|\bra{\rightarrow}\!\uparrow\rangle|^2$,'' which also happens to equal $\frac12$. Unlike our answer, these other answers are based only on intuitions, not a rigorous method of counterfactual reasoning.

\begin{figure}[h]
    \centering
    \includegraphics[width=0.5\textwidth]{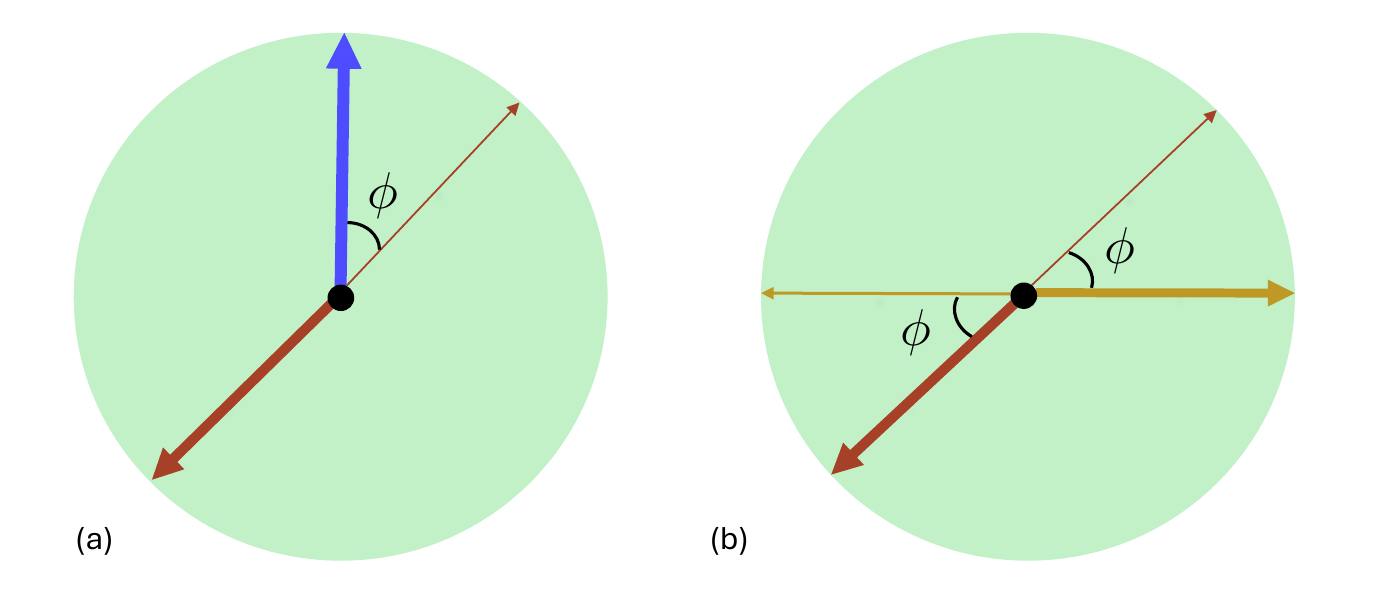 }
        \caption{States on the Bloch ball, with colours  following from Fig.\ref{fig:sidebyside}.   (a) Actual world: Alice's outcome $\uparrow$ and Bob's possible outcomes (probabilities $\approx0.15$ and $\approx0.85$) using his basis $Y=$\rotatebox[origin=c]{135}{$\updownarrow$} in the actual world.(b) Counterfactual world: Fixing Bob's outcomes, Alice's  possible outcomes in the basis $X'=\leftrightarrow$ (probabilities $=0.75$ and $=0.25$) in the counterfactual world. Note that the arrows have thickness corresponding to their actual or counterfactual probabilities, with the thickness of $\uparrow$ corresponding to probability one.}
    \label{fig:CHSHexplanation}
\end{figure}

%The result from \ref{eq:final} shows that Alice can infer a better estimate about the alternative scenario if she uses her knowledge of the actual outcomes in comparison to if she had directly measured on the singlet state.
%\subsection{Evaluating \erf{eq:final} \label{sec:Eq.3}}
The result $\frac{3}{4}$ for the counterfactual probability \erf{eq:final} can also be understood from 
Fig.~\ref{fig:CHSHexplanation}. Once Alice makes a measurement of $X=\updownarrow$ and obtains her outcome as $A=\uparrow$, this collapses Bob's state to $\ket{\downarrow}$. He measures with his chosen basis $Y=$\rotatebox[origin=c]{135}{$\updownarrow$}, making a Bloch-sphere angle of $\phi=\pi/4$ with Alice's measurement axis. Since Bob's outcomes are unknown to Alice, she can only estimate the likelihood of Bob's outcomes. Given Bob's collapsed state is $\ket{\downarrow}$, this determines the weights of his outcomes: $\cos^2(\phi/2)\approx0.85$ for \rotatebox[origin=c]{135}{$\uparrow$} (the one closer to $\downarrow$) and $\sin^2(\phi/2)\approx0.15$ for \rotatebox[origin=c]{135}{$\downarrow$}. Bob's possible outcomes are the fixtures linking the actual and the counterfactual world. Due to perfect anti-correlation of the shared singlet state, Alice can reason that in the counterfactual world she gets states  $|${\rotatebox[origin=c]{135}{$\downarrow$}}$\rangle$ and $|${\rotatebox[origin=c]{135}{$\uparrow$}}$\rangle$ with probabilities $\cos^2(\phi/2)$ and $\sin^2(\phi/2)$ respectively. Thus the inferred probability for her to get outcome $\rightarrow$, the supposability $\Su(A'=\rightarrow | X'=\leftrightarrow || A=\uparrow,X=\updownarrow,Y=\rotatebox[origin=c]{135}{$\updownarrow$})$, is given by $\cos^4(\phi/2)+\sin^4(\phi/2) = \frac{3}{4}$.
A more detailed explanation with evaluation of the probabilities explicitly can be found in~\cref{App:A}.

\section{The general counterfactual setting calculus \label{sec5:Calculus}} 
Now we formalize the rules to compute a supposability for counterfactual settings in any scenario. 
%specifying the way to identify the essential components that constitute a counterfactual and thus apply Lewis' analysis to any probabilistic theory. 
We begin %consider a counterfactual setting scenario 
by defining the set of all experimentally relevant events, $\mathbf{\Omega}=\{(\Omega_\alpha,\alpha)\}_\alpha$. Here $\alpha$ is a unique index for an event, including its space-time location, and $\Omega_\alpha$ is the classical random variable describing it. (Not all variables need be random, but we allow it.) 
This set contains a subset  
$\boldsymbol{Z}$ of measurement settings, 
and a disjoint subset $\boldsymbol{O}$ of corresponding outcomes (so $|\boldsymbol{Z}|=|\boldsymbol{O}|$). %specified by the subscript $\alpha$ and  
The remaining disjoint subset $\mathbf{\Lambda}$ comprises exogenous independent classical noise variables. Next, we need a physical theory defining probability distributions for the outcomes given the settings and noise, $\wp(\boldsymbol{o}|\boldsymbol{z},\boldsymbol{\lambda})$, and for the noises themselves,  $\mathbf{p}=(\ldots,p_l(\lambda_l),\ldots)$. 
We also need a causal structure $\frak{C}$ (such as a quantum DAG, or just space-time relations) on $\cu{\alpha}$, which will be reflected in the no-signalling properties of $\wp$. 
For the setting choice, we allow an arbitrary 
(even adaptive~\cite{wiseman2009quantum}) strategy $\boldsymbol{\mathcal{S}}$, a separate set of non deterministic functions (also constrained by $\frak{C}$) that specify each setting by $Z_j=\mathcal{S}_j(\boldsymbol{\Omega})$ (requiring $|\boldsymbol{\mathcal{S}}|=|\boldsymbol{Z}|$).

To define the counterfactual, we need the following. There is some evidence, $\boldsymbol{E} \subseteq \boldsymbol{Z} \union \boldsymbol{O}$ with value $\boldsymbol{e}$. A subset $\boldsymbol{A} \subseteq \boldsymbol{Z}$, of settings, defines the counterfactual antecedent, $\boldsymbol{A}'=\boldsymbol{a}'$. That is, the counterfactual strategy is $\boldsymbol{\mathcal{S}}'=\overline{\boldsymbol{{\mathcal{S}}}} \,\union\,\boldsymbol{a}'$, where $\overline{\boldsymbol{{\mathcal{S}}}}$ is a subset of $\boldsymbol{\mathcal{S}}$ 
which is disjoint to %$ =\boldsymbol{\mathcal{S}} \, \backslash \, 
$\boldsymbol{a}'$, where the latter is here being treated as a set of trivial functions. Define also $\boldsymbol{D}\subset\boldsymbol{\Omega}$, the inclusive
descendants, relative to the causal structure ${\frak{C}}$, of $\boldsymbol{A}$, and $\boldsymbol{D}'$ similarly for $\boldsymbol{A}'$. 
This lets us define the fixtures, the set of events such that $\boldsymbol{F}\equiv\boldsymbol{F}'$, by 
%\equiv\text{InclusiveDescendants}(\boldsymbol{A'})$.
%Lewis' second desiderata suggests that we  maintain perfect match of particular fact between the actual and the counterfactual world without violating any laws of the theory. This defines the fixtures between both the worlds and from the definition of $\boldsymbol{D}$ we can rewrite the list of events  $\boldsymbol{\Omega}=\boldsymbol{D}\cup\boldsymbol{F}$ where 
$\boldsymbol{F}\equiv\boldsymbol{\Omega}\backslash\boldsymbol{D}$. 
%The events in the subset $\boldsymbol{F}$ are outside any influence of the counterfactual antecedent and we can rewrite the events in counterfactual scenario as,
%    \begin{equation}       \boldsymbol{\Omega}'=\boldsymbol{D}'\cup\boldsymbol{F}'=\boldsymbol{D}'\cup\boldsymbol{F}\,.
%    \end{equation}
Finally, there is some consequent $\boldsymbol{C}'\subseteq\boldsymbol{D}'$, 
for which we can consider any possible value, $\boldsymbol{C}'=\boldsymbol{c}'$.

The counterfactual question can now be stated. With, as background, $\wp$, $\mathbf{p}$, and 
$\boldsymbol{\mathcal{S}}$, what is the 
supposability that $\boldsymbol{C}'$ would equal $\boldsymbol{c}'$, had $\boldsymbol{A}'$ taken value  $\boldsymbol{{a}}'$, given $\boldsymbol{E} = \boldsymbol{e}$?
As above, we notate and define this as
\begin{align}
&\Su(\boldsymbol{C}'=\boldsymbol{c}'|\boldsymbol{A}'=\boldsymbol{a}'||\boldsymbol{E}=\boldsymbol{e}) \coloneqq \label{countefactualcalculus} \\ 
&\sum_{\boldsymbol{f}} \text{Pr}(\boldsymbol{C}'=\boldsymbol{c}'|\boldsymbol{A}'=\boldsymbol{a}',\boldsymbol{F}=\boldsymbol{f})\,\text{Pr}(\boldsymbol{F}=\boldsymbol{f}|\boldsymbol{E}=\boldsymbol{e}). \nonumber 
\end{align}
Note that the second Pr is evaluated (using  $\wp$ and ${\bf p}$) for strategy  $\boldsymbol{\mathcal{S}}$, while the first 
is for strategy  $\boldsymbol{\mathcal{S}}'=\overline{\boldsymbol{{\mathcal{S}}}} \,\union\,\boldsymbol{a}'$. %\boldsymbol{\overline{\mathcal{S}}}$ 

%\sechead{Comparison with another approach} 

%Finally, for the consequent $\boldsymbol{C'}=\textbf{c}'$, they consider the complement of $\boldsymbol{B'}$. i.e., $\boldsymbol{\textbf{A}'=(B',C')}$. 
\subsection{Comparison with other Work \label{Sec:Otherworks}}
As noted in~\cref{sec3:Lewis}, a recent work, by Suresh \ea~\cite{suresh2023semantics}, previously introduced counterfactual probabilities in a similar context to our own work. %They based their ``semantics for counterfactuals'' on 
% Most critically, in their semantics, the only thing fixed across the actual and counterfactual worlds is the classical noise $\boldsymbol{\Lambda}$. Thus, in scenarios (such as the CHSH example above, and the one to follow) with no classical noise, they would obtain only trivial counterfactual probabilities, in which the evidence from the actual world is irrelevant. \ing{From here} 
The crucial points of difference with our work start with the fact that they take inspiration from Pearl~\cite{pearl2000models} rather than Lewis. The former uses a classical causal model, represented by a DAG, which was generalized to a quantum causal model in~\cite{suresh2023semantics}. 
%In our case, we do not commit to any particular quantum DAG. 
Following Pearl, their formalism proceeds in three main steps: abduction, action, and prediction. 

In the abduction step, Bayesian inference is used to update the probabilities of the unknown (exogenous) variables given the evidence. For us, the evidence consists of the all the settings and only a subset of the outcomes, while they `canonically' take the evidence to be {\em all} results and {\em all} settings (in our notation, $\boldsymbol{e}=(\boldsymbol{a},\boldsymbol{z})$). Also, their semantics does not allow adaptive measurement schemes. A more crucial difference is that the only variables they perform this Bayesian updating on are the classical noise variables $\boldsymbol{\Lambda}$. Thus, in the scenarios we consider in this Letter, with no classical noise, they would obtain only trivial counterfactual probabilities, in which the evidence from the actual world is irrelevant. 

In the action step of Pearl's formalism, the antecedent of the counterfactual is determined by actively intervening in the DAG and fixing the value of the corresponding random variable, thereby severing all causal arrows that lead to that variable. Their antecedent includes {\em all} the measurement settings $\boldsymbol{z}'$ and a sublist of the corresponding outcomes $\boldsymbol{b}'$ obtained at a set of nodes $\boldsymbol{B}$. It is this choice of antecedent that entails that only the exogenous variables $\boldsymbol{\Lambda}$ are fixed across the actual and counterfactual worlds, because all other variables are descendents of $\boldsymbol{z}'$ in the DAG. Lewis, by contrast, avoided this kind of active-intervention step, treating the antecedent as a plausible alternative within the laws of the theory. This is most easily satisfied by restricting to settings for the antecedent as we have done --- meaning our formalism is strictly different from, rather than more general than, the semantics of Ref.~\cite{suresh2023semantics} --- but it is not out of the question to also consider outcomes as antecedents in Lewis's framework, as noted in our conclusion.

%
%In our formalism, it is fairly straightforward to refer to a different measurement choice without violating any laws of QT, but we do not have any outcomes in the antecedent. 
The third and final step in Pearl, and in Ref.~\cite{suresh2023semantics}, is prediction, where the antecedent and the updated information about the classical noise $\boldsymbol{\Lambda}$ is used to compute the counterfactual probabilities of the measurement outcomes $\boldsymbol{c}'$ from nodes $\boldsymbol{C}$. Here $\boldsymbol{C}$ is the set of outcome nodes that are disjoint from those ($\boldsymbol{B}$) forming part of the antecedent. This step is analogous to our evaluation of the supposabilities of consequents (also denoted $\boldsymbol{C}$).
%different possible outcomes). Their consequent is the outcomes corresponding to the node  denoted as $\textbf{c}'$. 

%\ing{Discussion on the paper on joint exogeneity}.
Very recently, another interesting paper considering counterfactuals in quantum mechanics has appeared~\cite{wang2025quantumexperimentjointexogeneity}. This one is more in the spirit of counterfactual definiteness as discussed in the Introduction, but  Wang and Zhang consider the novel assumption of joint exogeneity (that the collection of all potential outcomes in any randomized experiment is statistically independent of the treatment assignment). They show that this assumption, interpreted in a particular way, can be falsified by OQT in the so-called instrumental scenario. Here we wish to make clear that the counterfactual question we ask does not rely on any joint exogeneity assumption, even implicitly. We do not postulate a joint distribution over the outcomes of different measurements of Alice. Critically, our supposability (not a probability) of the  consequent has a  type of conditioning ($||$) on events in the actual world that is different from its conditioning ($|$) on those in the counterfactual world. 
%but rather, we define a conditional of the counterfactual on the actual via our fixed events (Bob's settings and outcomes). This conditioning is not direct and also conceptually different from the conditioning that is taken in probability theory. On expansion this conditioning reduces to normal conditional probabilities meditated via Bob's distributions which are fixed in both the scenarios.

\section{Counterfactuals with continuous monitoring \label{sec6:}} Here we show the power of our formalism by applying it in a scenario with two parties continuously monitoring a quantum system. 

\subsection{Description of the System} 
\begin{figure}
    \centering
    \includegraphics[width=0.45\textwidth]{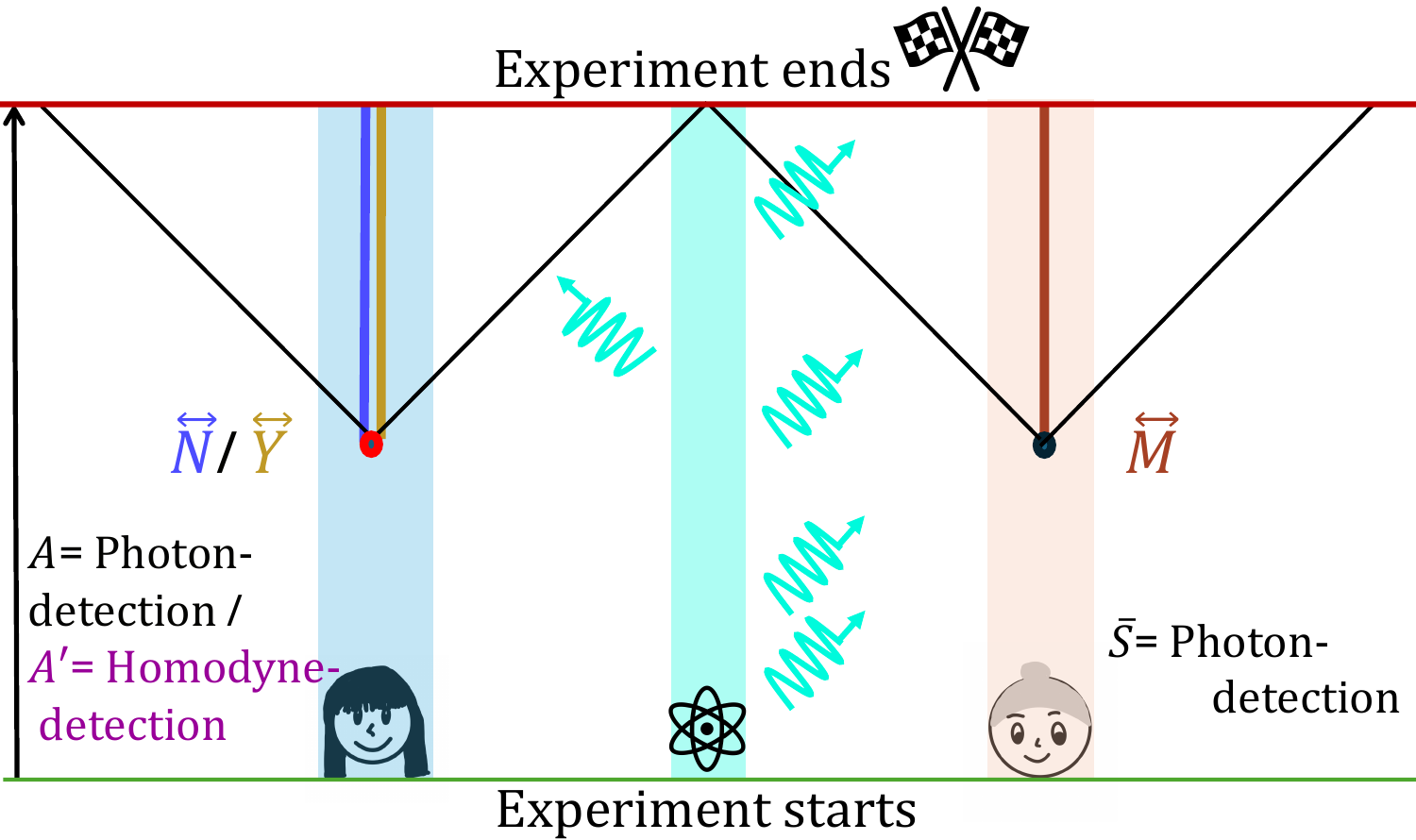 }
        \caption{Space-time diagram %corresponding to Fig.~\ref{fig:atom in a cavity}. 
        for the fluorescence counterfactual. Alice's actual record of photo-detections $\overleftrightarrow{N}$ (blue, evidence) is space-like separated from Bob's record of photo-detections $\overleftrightarrow{M}$ (brown, fixture). Alice's counterfactual measurement is homodyne, with record $\overleftrightarrow{Y}$ (golden, consequent). The photons (cyan wiggles, not classical events but included to aid intuition) are emitted by the atom asymmetrically in accordance with the chosen parameters for the simulation: $\eta_A=0.2$, $\eta_B=0.8$, and $\Omega=2\gamma$, for an interval $[0,10\gamma\inv)$. }
    \label{fig:SpaceTimediagram}
\end{figure}

A resonantly driven two-level atom radiates, say via a leaky cavity, into two independent electromagnetic baths, % (or environments) 
which separately undergo continuous-in-time measurements by two observers, Alice and Bob, as in Fig.~\ref{fig:SpaceTimediagram}. 
%{fig:atom in a cavity}.
%\begin{figure}[H]
%    \centering
%    \includegraphics[width=0.45\textwidth]{Images/Continuous scheme.jpg }
%        \caption{A counterfactual with continuous-in-time measurements: Alice and Bob continuously monitor an atom via their respective photon detectors, yielding records $N$ and $M$. Alice's counterfactual question (green thought bubble) is: given her photon-count record $N$, what would she have observed if she had measured using a homodyne detection instead? The colour scheme for various elements follows that of Fig.~\ref{fig:sidebyside}. }
%    \label{fig:atom in a cavity}
%\end{figure}
In the interaction frame, the atom's unconditioned state obeys the Lindblad master equation 
\begin{align}
    \rho_{t+\dd t}&=(1+\mathcal{L}\dd t)\rho_t \nonumber\\ & 
    = \gamma \dd t \mathcal{J}[\hat{\sigma}_-]\rho_t+\mathcal{J}[(1-i\hat{H}_{\text{eff}}\,\dd t)]\rho_t , \label{eq:Lindblad}
\end{align}
where $\mathcal{J}[\hat{c}]\bullet\equiv\hat{c}\bullet\hat{c}^\dagger$, and %$\hat{H}_{\text{eff}}=\frac{\Omega}{2}\hat{\sigma}_x-i\frac{\gamma{2}\hat{\sigma}_+\hat{\sigma}_-$
$2\hat{H}_{\text{eff}}=\Omega\hat{\sigma}_x-i\gamma\hat{\sigma}_+\hat{\sigma}_-$~\cite{wiseman2009quantum}. 
Here $\Omega$ is the Rabi frequency and $\gamma$ the total radiative damping rate (from both ends of the cavity).
%Radiative damping is described by a master equation with the Lindblad operator $\sqrt{\gamma\eta}\hat{\sigma}_-$~\cite{wiseman2009quantum}. 

We assume that a fraction $\eta_A$ of the fluorescence is detected by Alice, while Bob detects the rest: $\eta_B = 1 - \eta_A$. In the actual world, both Alice and Bob monitor the fluorescence through photon detection, resulting in all-time records denoted by $\overleftrightarrow{N}$ and $\overleftrightarrow{M}$ respectively, in the notation of 
Ref.~\cite{PhysRevLett.115.180407}. To understand continuous measurement theory, concentrate on Alice. Her record $\past{N}_t$ (the record up to time $t$) refines her knowledge of the quantum state from $\rho_t$ to the conditional state ${\rho}^{\past{N}}_t$, which obeys~\cite{wiseman2009quantum} 
\begin{equation}
\tilde{\rho}^{\past{N}}_{t+\dd t}=
\begin{cases}
     \gamma\eta_A \dd t \mathcal{J}[\hat{\sigma}_-]\tilde{\rho}^{\past{N}}_t  \text{ if her detector clicks},\\ \{1+(\mathcal{L}-\gamma\eta_A\mathcal{J}[\hat{\sigma}_-])\dd t\} \tilde{\rho}^{\past{N}}_t  \text{ otherwise},
\end{cases}     \label{jumptraj}
\end{equation}
Here, for convenience, we are giving the equations for the state $\tilde{\rho}$ whose norm is the probability of the record. The time-argument for the $\past{N}$ on the left-hand-side is $t+\dd t$.

As per Fig.~\ref{fig:SpaceTimediagram}, %{fig:atom in a cavity}, 
Alice is considering the counterfactual world where she measures a field quadrature instead, by a $Y$-homodyne detection, which would yield a record $\both{Y}$. The analogous 
conditioning equation to \erf{jumptraj} is~\cite{wiseman2009quantum}
\begin{equation}
\dd \tilde{\rho}_t^{\overleftarrow{Y}}=\dd t\mathcal{L}\tilde{\rho}^{\overleftarrow{Y}}_t+\sqrt{\eta_A\gamma}\;\dd W_t[i\hat{\sigma}_{-}\tilde{\rho}^{\overleftarrow{Y}}_t+\tilde{\rho}^{\overleftarrow{Y}}_t(-i\hat{\sigma}_+)]\,, \label{diffusion}
\end{equation}
where $\dd W_t$ is Wiener noise, which is related to the photocurrent (with a convenient choice of normalisation) by
\begin{equation}
    Y_t=\text{Tr}[\rho^{\overleftarrow{Y}}_t\hat{\sigma}_y]+\frac{1}{\sqrt{\eta_A\gamma}}\frac{\dd W_t}{\dd t}. \label{homodynecurrent}
\end{equation}

\subsection{The Counterfactual Question}
Consider this counterfactual question by Alice: 
``Given that Bob and I are both using photon counters, and I detected the photon click record $\both{N}=\overleftrightarrow{n}$ in interval $[0,T)$, what is the supposability for  $\overleftrightarrow{Y}$ in the same interval, had I used a Y-homodyne detector instead?'' 
For continuous measurements, even in a finite time $[0,T)$, the cardinality of  $\boldsymbol{Z}$, $\boldsymbol{O}$, and $\boldsymbol{\mathcal{S}}$ is $\aleph_1$. However, in our scenario, the measurement settings are constant in time, in both the actual and counterfactual worlds. Thus, rather than attempt explicit set representations, we will use a descriptive notation 
that we trust the reader will understand. 
Splitting things by party (Alice; Bob), we have %can elide the difference between $\boldsymbol{Z}$ and $\boldsymbol{\mathcal{S}}$, and  
\begin{align}
    \boldsymbol{\mathcal{S}} & =\text{(photon-counting; photon-counting)},\label{ctsS} \\
  \boldsymbol{\mathcal{S}}' & = (\boldsymbol{a}'; \overline{\boldsymbol{\mathcal{S}}}) =\text{(Y-homodyne; photon-counting)}. \label{ctsSp}
\end{align}
As in the CHSH case, the evidence $\boldsymbol{E}$ is Alice's setting and her results, and Bob's setting, and here has the value  
$
\boldsymbol{e} =\text{(photon-counting, }\both{n} ; \text{photon-counting)}$.

Now, to evaluate the counterfactual, we need to consider the space-time diagram in Fig.~\ref{fig:SpaceTimediagram}. 
%for the continuous monitoring scenario justifies the choice of the fixed events. 
The antecedent $\boldsymbol{A}$ is all of Alice's measurement settings through time. In the time interval of the experiment, the only events in the future light cone of these are her outcomes. Thus $\boldsymbol{D}$, the inclusive descendants of the antecedent, are her settings and outcomes. Therefore the fixture $\boldsymbol{F}$ comprises the complement to this, namely 
Bob's settings and outcomes: 
%\beq 
$\boldsymbol{F} \equiv (\nullset ; \text{photon-counting, }\both{M})$.
%\eeq

Finally,  
the consequent is Alice's counterfactual homodyne current record, $\boldsymbol{C}'\equiv\overleftrightarrow{Y}$. Now we are confronted with the cardinality problem: to calculate the full supposability distribution for this nowhere-differentiable function of time is infeasible. Therefore we focus on a simple statistic for it:  the {\em suspectation} of $Y_t$ at each $t\in[0,T)$.  Here we have introduced suspectation ($\mathbb{S}$) to be the counterfactual counterpart to expectation ($\mathbb{E}$), just as supposability is to probability.  
That is, suppressing the information that is contained in the given $\boldsymbol{\mathcal{S}}$ (\ref{ctsS}), 
and using similar notation to \erf{countefactualcalculus}, we want 
\begin{align}
&\mathbb{S}[Y_t|\boldsymbol{A}'=\boldsymbol{a}'||\both{N}=\both{n}] \nn \\
&\coloneqq \int \dd y \, 
y \,  \dbd{y}\Su[Y_t \leq  
%\Su[Y_t \geq
y|\boldsymbol{A}'=\boldsymbol{a}'||\both{N}=\both{n}] \label{defsuspect}
\end{align}
(recall $\boldsymbol{a}'=$ Y-homodyne). 
By \erf{countefactualcalculus}, this equals
\begin{align}
\sum_{\both{m}} \mathbb{E}[Y_t|\boldsymbol{A}'=\boldsymbol{a}',\both{M}=\both{m}]
 \Pr(\both{M}=\both{m}|\both{N}=\both{n}). \label{twotermsinsuspect}
\end{align} 

%(The notation $\leftrightarrow$ implies the full-time record for the interval $[t_i,t_f)$\cite{guevara2015quantum}). 

\subsection{Computation of the Suspectation Value \label{sec:EM1}}

Numerically evaluating \erf{twotermsinsuspect} requires calculating a regular expectation value ($\mathbb{E}$) and averaging it over 
all values of $\both{m}$ with the appropriate probability (Pr). 

For performing the average, we use the method developed for quantum state smoothing~\cite{PhysRevLett.115.180407,chantasri2021unifying}. This uses the probability-normalized conditional state evolution introduced above; see~\cref{App:C4} for the details in our case. 

For the expectation value, we can apply the theory of generalized quantum weak values~\cite{PhysRevA.65.032111,chantasri2021unifying,PhysRevLett.111.160401}. This is applicable because homodyne measurement in a time interval $[t,t+\dd t)$ of infinitesimal duration is a weak measurement, and because we are considering the value conditioned on the results of other measurements ($\both{M}$) in both the past ($\past{M}$) and future ($\fut{M}$) of the time $t$ of the homodyne measurement. The weakness of the measurement can be seen from \erf{homodynecurrent}, where the signal (first term) is contaminated by noise (second term) whose variance diverges. The role of a particular past measurement record $\past{m}$ is described by its preparation of a conditioned state $\rho^{\overleftarrow{m}}_t$ at time $t$, 
while the role of a particular future measurement record $\fut{m}$ is described by its ``post-paration'' of a POVM element or effect  $\hat E^{\overrightarrow{m}}_t$~\cite{1983LNP...190.....K,wiseman2009quantum,Helstrom:1969fri} which applies at time $t+\dd t$. The former obeys (giving the unnormalized version for convenience) 
\begin{equation}{\label{S10}}
    \tilde{\rho}^{\past{M}}_{t+\dd t} = 
\begin{cases}
\cu{1 + \dd t\ro{ {\cal L} + \lambda - \eta_{\rm B}\gamma{\cal J}[\sigma_{-}]}}\tilde{\rho}^{\past{M}}_t  & \text{if no click,}\\
  (\eta_{\rm B}\gamma/\lambda){\cal J}[\sigma_{-}] \tilde{\rho}^{\past{M}}_t & \text{if click.}
   
  \end{cases} 
\end{equation}
with $\tilde{\rho}_0=\rho(t_0)$. 
For the latter, $\hat{E}_T=\hat I$ and $\hat E^{\fut{M}}_{t-\dd t}$ is related to $\hat E^{\fut{M}}_{t}$ by the {\em adjoint} of the maps in \erf{S10}. 

Homodyne measurement differs from the usual weak measurement considered in weak values~\cite{PhysRevLett.60.1351} because the coupling of the atom to the bath (here playing the role of the meter) is dissipative, meaning there is extra back-action. This is manifested in the weak value formula involving the real part of a weak-value of the atomic lowering operator $-i\s{-}$[as appearing to the left of the state in \erf{diffusion}] rather than the Hermitian part of that operator, $\s{y}$ [as appearing in \erf{homodynecurrent}]. Specifically, the pre- and post-selected mean homodyne current at time $t$ is 
\beq 
\mathbb{E}[\text{Y}_t|\overleftarrow{M}=\past{m},\overrightarrow{M}=\fut{m}] = {\rm Re}\frac{\Tr[\hat E^{\overrightarrow{m}}_t (i\hat{\sigma}_{-})\rho^{\overleftarrow{m}}_t ]}{\Tr[\hat E^{\overrightarrow{m}}_t \rho^{\overleftarrow{m}}_t]} .
\eeq
\subsection{Results}

We consider the case where Alice's actual record is a single click in her detector, at time $t_A=6.25 \gamma\inv$. This is a typical record for the parameters we choose, $\eta_A=0.2$, and $\Omega=2\gamma$, for an interval $[0,10\gamma\inv)$.  We choose this record so that we can see clearly the interesting features in the suspectation of $Y_t$ around this time. 
\begin{figure}
    \centering
    \includegraphics[width=0.50\textwidth]{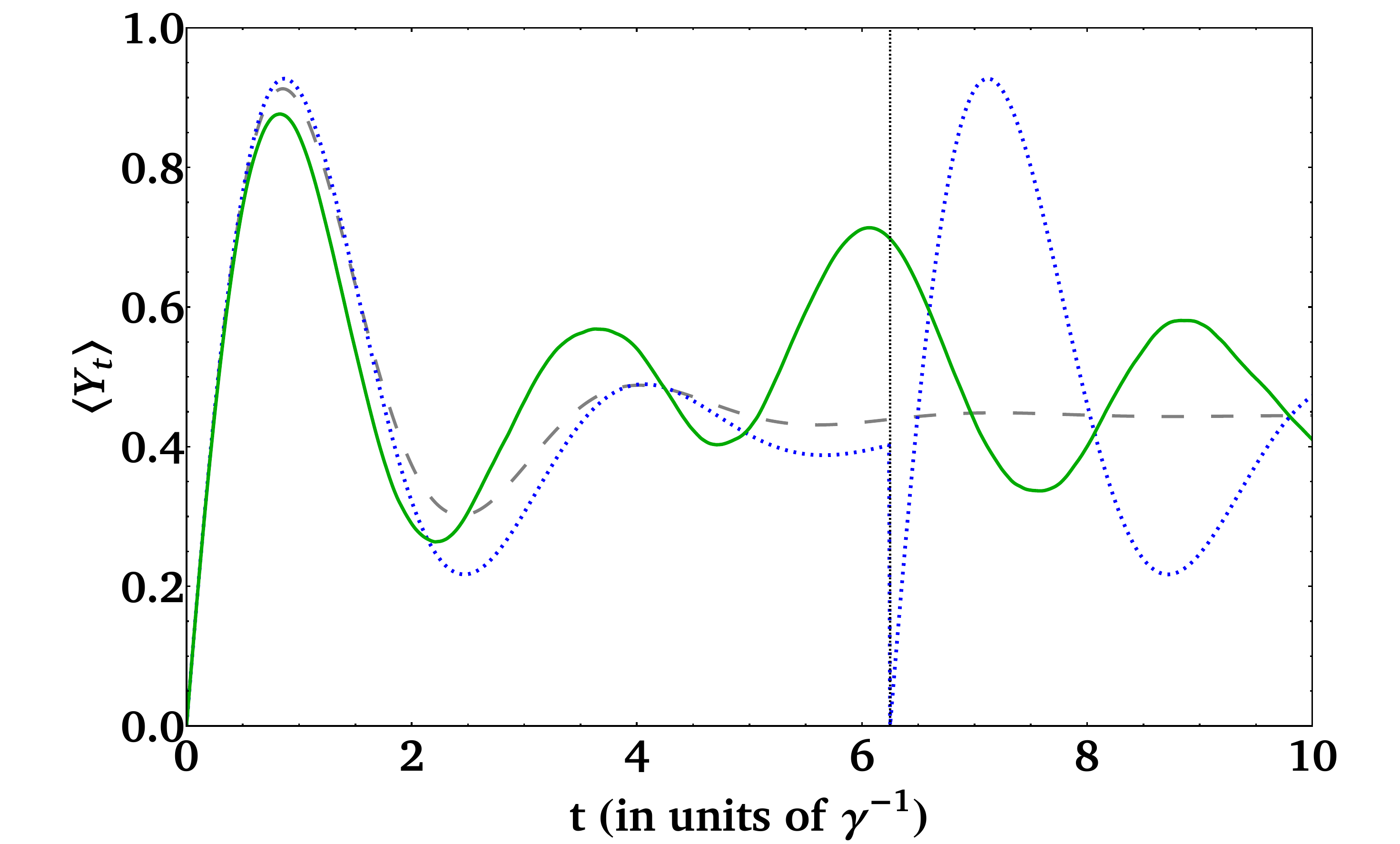 }
    \caption{\small{Dynamics of various estimates by the Alice of Fig.~\ref{fig:SpaceTimediagram}. The expectation of $\s{y}$ is shown for the unconditioned atomic state (gray dashed line) and Alice's filtered state conditioned on her actual record (a photo-detection at $t_A=6.25\gamma\inv$) (blue dotted line). 
    The gray dashed line is also the unconditioned expectation ${\mathbb E}[Y_t]$ of Alice's Y-homodyne current (\ref{homodynecurrent}). Finally, the green solid curve
    curve is the suspectation ${\mathbb S}[Y_t]$ for (\ref{homodynecurrent}), conditioned on her actual record $\both{n}$, as in \erf{defsuspect}. 
    }}
    \label{fig:Results-Hom}
\end{figure}

The results are given in Fig.~\ref{fig:Results-Hom}. The gray line is the evolution of $\Tr[\rho_t\s{y}]$, with 
$\rho_t$ obeying the Lindblad equation (\ref{eq:Lindblad}), starting from the ground state. From \erf{homodynecurrent}, this is also the expectation value of the homodyne photocurrent, if it were being performed. The blue line shows the same for Alice's filtered state $\rho_t^{\past{n}}$, obeying \erf{jumptraj}, for the specific record of one click at time $t_A=6.25 \gamma\inv$. At that time, the atom jumps back to the ground state, where $\Tr[\rho_t\s{y}]=0$. In this case, the homodyne detection is not actually being performed, obviously. The green line shows \erf{defsuspect}, the suspectation of Alice's counterfactual homodyne current, \erf{homodynecurrent}, given the evidence of her actual photon count record. Note that this 
%The filtered results of her actual photo detection records show that given that the atom is initially in the ground state it begins the rabi-oscillations when the drive is turned on and when the detector clicks at time $t=6.25$, it restarts the rabi-oscillations from the ground state as was in the initial time.
%The 
suspected value 
peaks around the time of the actual jump, and the ``echoes'' of this peak are seen in the $\Omega$-frequency oscillation of the value both before and after it. We now turn to explaining why this peak arises.

%UP TO HERE This reflects the fact that Alice actually has recorded a photon at that time, which contributes to the intensity of the counterfactual homodyne current conditioned on the actual events. A more detailed analysis of the results has been given in the supplementary material~\cite{SM}.
\subsection{Understanding the Suspectation Value \label{sec:EM2}}%curve shape in Fig.~\ref{fig:Results-Hom} \label{sec:EM2}}

In the actual world, where $\boldsymbol{O}\equiv(\both{N};\both{M})$ and $\both{N}=\both{n}$, the anti-bunching of resonance fluorescence~\cite{HJCarmichael_1976} means Alice can infer that 
Bob's rate of clicks is suppressed in the time either side of Alice's actual jump (at $t_A=6.25\gamma\inv$). It is these clicks (only a handful over the whole time interval $[0,10\gamma\inv)$ in any individual run) that are fixed in the counterfactual world. The exact times of Bob's clicks are unknown to Alice, but she knows that their rate is suppressed at time near $t_A$. 

In the counterfactual world, where $\boldsymbol{O}'\equiv(\both{Y};\both{M})$, the suppression of Bob's jumps can only be due to Alice's counterfactual record $\both{Y}$ causing  
%But what feature of $\both{Y}$ would suppress Bob's click rate near $t=6.25\gamma\inv$? 
%Bob's click rate will be suppressed (relative to the unconditioned rate) if his 
the conditioned atomic state to be closer to the ground state then than it would otherwise be. Thus, we need to understand for what type of 
homodyne record $\both{Y}$ will the dynamics of \erf{diffusion}  increase the ground-state population 
$\rho^{\past{Y}}_{gg} \equiv \langle g|\rho^{\past{Y}}|g\rangle$ of the atom near $t_A=6.25\gamma\inv$. 

\begin{figure}[H]
    \centering
    \includegraphics[width=0.50\textwidth]{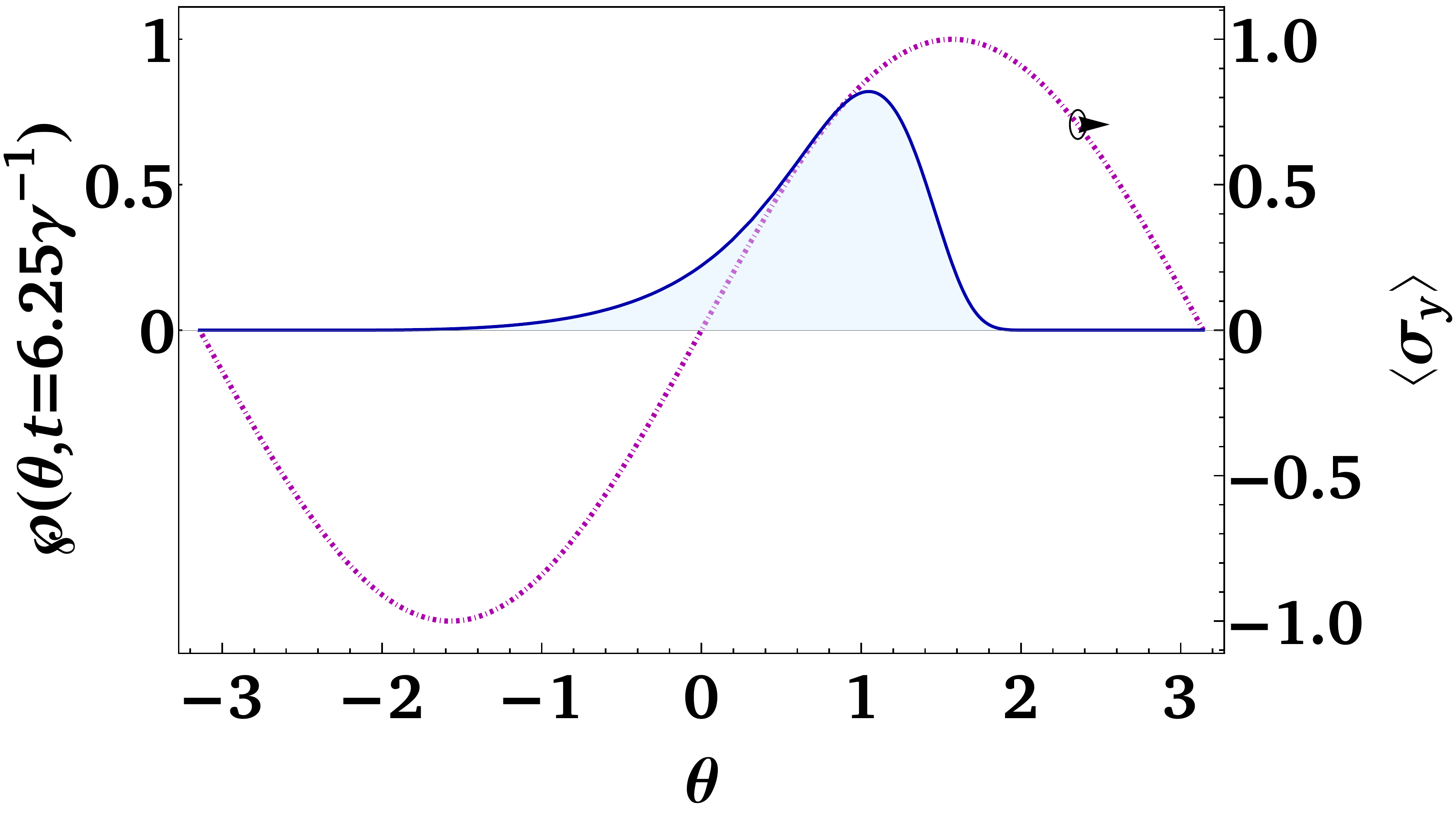}
    \caption{Tendency of $\langle\hat\sigma_y\rangle$ to be positive. The blue curve, with light blue filling, is the normalized probability distribution of the conditioned state ${\rho}_t^{\past{M},\past{Y}}$ at time $t=t_A$. This state lives on the $y$--$z$ plane of the Bloch sphere and is thus parametrized by the single variable $\theta$. It is  
    conditioned on both Alice's Y-homodyne measurement ($\past{Y}$) and Bob's photo-counts ($\past{M}$). We take $\past{M}$ to be a fixed $\past{m}$, with a jump at time $t=4.71\gamma\inv = t_A - 1.54\gamma\inv$, and no jumps in the interval $(4.71\gamma\inv,t_A)$. This is the most  characteristic record for Bob ($\past{m}^\chi$) since the rate of his jumps, conditioned on $\both{N}=\delta_{t,t_A}$, has a maximum $\approx 0.6\gamma$ at $t=4.71\gamma\inv$; see~\cref{App:C5}. The total probability (from integrating the curve) in the region $\theta > 0$ is $\approx 90 \% $. This corresponds to the region where $\langle\hat\sigma_y\rangle > 0$, as seen from the magenta dot-dashed curve which shows the sinusoidal curve which is $\langle\hat\sigma_y\rangle$ as a function of $\theta$. 
    \label{fig:Positive Y}}
\end{figure}

By normalizing \erf{diffusion}, it is easy to show that 
\begin{equation}
    {\dd} \rho^{\past{Y}}_{gg} \propto \langle \hat{\sigma}_y\rangle^{\past{Y}}\ro{1-\rho^{\past{Y}}_{gg}}{\dd}W_t+O(\dd t)\,, \label{changegroundpop}
\end{equation}
where ${\dd}W_t = \ro{Y_t - \langle \hat{\sigma}_y\rangle^{\past{Y}}}\dd t$ is the difference between the current and its expected value from filtering alone [see \erf{homodynecurrent}]. Thus, if the atomic ground state population is to increase, the right-hand-side of \erf{changegroundpop} must tend to be positive. The term inside the bracket is always positive, and it turns out that the 
expectation value of $\hat{\sigma}_y$ is overwhelmingly likely to be positive at $t=t_A$ (see Fig.~\ref{fig:Positive Y}), when considering a typical Bob record.
Thus, for Bob's clicks to be suppressed near $t_A=6.25\gamma\inv$, it must be the case that $\dd W_t$ tends to be positive in that vicinity. This causes $Y_t$ to peak at around that time, exactly as seen in Fig.~\ref{fig:Results-Hom}. Since $Y_t$ is generated by a dynamical process [Eqs.~(\ref{diffusion}) and (\ref{homodynecurrent})], the selection of a peak around $t_A=6.25\gamma\inv$ entails the appearance of 
%and we are only looking at a specific time window it causes the 
ripples at earlier and later times.  
%to adjust within the given frame Once the conditioning on the evidence (the detected photon) is removed it will simply be the current due to master equation solution.

    \section{Conclusion \label{sec7:Conclusion}}
    We have introduced a general counterfactual calculus for non-deterministic theories,  replacing Lewis's truth-valued counterfactuals~\cite{lewis1979counterfactual} with probability-valued ones, which we call supposabilities. A key aspect of Lewis's approach we build upon is to maximize the fixtures across actual and counterfactual worlds. Our calculus is applicable to any probabilistic theory having free measurement choices with classical outcomes, and a causal (in the no-signalling sense) structure. In particular, we have applied it to quantum scenarios, from simple projective measurements on a Bell pair to continuous measurements of fields radiated by a driven atom.  As well as opening new directions in quantum causal modelling~\cite{brukner2014quantum} and quantum machine learning~\cite{QML}, our approach could lead to better grounded quantum generalizations of other theories of fundamental phenomena, such as Integrated Information Theory~\cite{albantakis2023integrated}, in which counterfactual reasoning is central. 

%We have proposed a less conservative approach compared to~\cite{suresh2023semantics} allowing a broader class of measurements (such as adaptive measurements) and allowing the unknown measurement results outside the causal future of the antecedent to be fixed. We have applied the formalism to projective measurements and to continuous-in-time measurements where in the later case we have shown how to answer questions pertaining to atomic fluorescence.  

In both of the examples we analysed, there are two parties with  no communication. Our calculus allows for scenarios with multiple parties and/or communication, which we will explore in future work. We can also generalize from calculating the suspected value of a photocurrent $Y_t$ at each time $t$ to calculating the suspectation of more interesting summary statistics of the whole record. A greater challenge %for future work 
is to relax the restriction that the antecedent of the counterfactual be a measurement setting. 
%The example in the continuous measurement scenario is strictly in two party correspondence where none of the outcomes of the spatially separated party can be causally influenced due to the antecedent. In our future work we aim to investigate a more general scenario where there might be potential  influence on the outcomes of Bob due to the change in settings by Alice which may arise due to overlap of the future light cones within the time frame of the experiment. We also aim to formalize the definitions of the elements where space like separation might not be the only causal connection. 
Finally,  the supposabilities we introduce can always be found empirically in the sense that all of the probabilities that enter the definition can be measured experimentally as relative frequencies; the question of whether they can also be measured more directly is open. 
%can the supposabilities in our theory be measured experimentally, beyond merely %$$  we would like extend this formalism for experimental implementation beyond 
%using empirical probabilities in place of theoretical ones? 

\emph{Note added.---} Following the posting of this paper on the arxiv, it was pointed out in Ref.~\cite{liu2025unifyingquantumsmoothingtheories} that the expected value of any observable $\hat{O}$ with respect to the smoothed quantum state~\cite{PhysRevLett.115.180407,chantasri2021unifying} at time $t$ is, in fact, the suspected value for a {\em counterfactual} measurement of $\hat O$ at time $t$ according to the theory introduced here. The smoothed quantum state also has an independent definition as the optimal estimate of a maximally conditioned quantum state~\cite{chantasri2021unifying}, and this interpretation of its expectation values has been verified experimentally in both optics~\cite{yokoyama2025trackingquantumdynamicsoptical} and optomechanics~\cite{khademi2025postprocessedestimationquantumstate}. Thus there is already at least one sense in which the final open question posed in the preceding paragraph can be answered positively.

\acknowledgements We thank  Y\`il\`e Y{\=\i}ng, Marwan Haddara, Pierre Guilmin, Ken Wharton, Robert Spekkens, Emily Adlam, Eric Cavalcanti, and Fabio Costa for discussions. This work was supported by the ARC Centre of Excellence project number CE170100012  and an Australian Government RTP Scholarship.
% \bibliographystyle{apsrev4-2}
%\bibstyle{unsrt}

% \hmw{Ingita, please fix references as per Kiarn's comments. Except I don't understand ``I suggest initials for all names excepts the last'', oh, unless Kiarn just means give the full surname. Yes, obviously.}\ktl{Yes that is what I meant.}\hmw{ Also, I don't know how Aspect became an author on [21]!}\ing{Fixed the referred reference.}
%\newpage
\appendix
\section{Counterfactual Probabilities for the Bell-CHSH scenario \label{App:A}}
\begin{table*}[t]
\centering
\begin{tabular}{ |p{4.5cm}|p{4cm}|p{6.5cm}|}
\hline
\multicolumn{3}{|c|}{Direct Mapping} \\
\hline
Terminology & Symbols & CHSH \\
\hline
Events     & $\mathbf{\Omega}$  & $\{X,Y,A,B\}$, $\mathbf{\Lambda}=\nullset$ \\

Measurement strategy         & $\boldsymbol{\mathcal{S}}$  & $X,Y$ \\
Measurement settings    & $\boldsymbol{Z=z}$  &  $\{X,Y\}=\{\updownarrow,\neswarrow\}$ \\
Measurement outcomes       & $\boldsymbol{O}$  & $\{A,B\}$ \\
Evidence        & $\boldsymbol{E}=\boldsymbol{e} $
%\,(\subseteq\boldsymbol{O=o\union Z=z})$  
& $\{A,X,Y\}=\{\uparrow,\updownarrow,\neswarrow\}$\\
Antecedent         & $\boldsymbol{A}'=\boldsymbol{a}'$ & $\{X'\}=\{\leftrightarrow\}$  \\
Counterfactual strategy & $\boldsymbol{\mathcal{S}}'(=\overline{\boldsymbol{\mathcal{S}}}\union \boldsymbol{a}')$  & $X',Y$ ($\overline{\boldsymbol{\mathcal{S}}}=Y$, which is the fixed strategy)\\
$\text{InclusiveDescendants}\,(\boldsymbol{A}')$ %($\boldsymbol{A}\subseteq\boldsymbol{Z}$ \hmw{what for?}) 
&  $\boldsymbol{D}'$ & $\{X',A'\}$ (Alice's settings and outcomes) \\
Consequent         & ${\bf C}'={\bf c}'$  & $\{A'\}=\{\rightarrow\}$ \\
Fixed events         & $\boldsymbol{F}$  & $\{B\}$ (Bob's outcomes) \\
Supposability       &  ${\rm Su}({\bf C}'= {\bf c}'|{\bf A}'={\bf a}'||{\bf E}={\bf e})$  & ${\rm Su}(A'=\,\rightarrow | X'=\,\leftrightarrow|| A=\uparrow,X=\,\updownarrow,Y=\neswarrow)$ \\
\hline
\end{tabular}
\caption{The first row is the list of terms that are used in our general formalism, the second row are the symbols used in the formalism corresponding to the terminology and the third row has the symbols used in the CHSH example corresponding to each terminology in the general formalism.}\label{tab:CHSH_mapping}
\end{table*}
Here, we provide a detailed, step-by-step explanation of how we evaluate the result $\frac{3}{4}$ for the counterfactual probability (supposability) in~\erf{eq:final}. 

We start by computing the individual probabilities occurring in the supposability

\begin{align}
{\rm Su}\!\left(A'=\rightarrow \mid X'=\leftrightarrow 
  \,\middle\|\, A=\uparrow, X=\updownarrow,
  Y=\rotatebox[origin=c]{135}{$\updownarrow$}\right)\nonumber
&\\= \sum_b 
\Pr\!\left(A'=\rightarrow \mid B=b, X'=\leftrightarrow,
  Y=\rotatebox[origin=c]{135}{$\updownarrow$}\right)
\nonumber
& \\\times
\Pr\!\left(B=b \mid A=\uparrow, X=\updownarrow,
  Y=\rotatebox[origin=c]{135}{$\updownarrow$}\right).
\label{Supp1}
\end{align}
Recall the actual scenario: Alice and Bob share a singlet state 
$\ket{\Psi_-}=\frac{1}{\sqrt{2}}(\ket{\uparrow\downarrow}-\ket{\downarrow\uparrow})$ and perform a Bell-CHSH experiment on it. The joint probability distribution of the outcomes given the settings is
\begin{equation}
    \wp(a,b|x,y)=\text{Pr}(A=a,B=b|X=x,Y=y)\,.
\end{equation}
Although Alice does not know Bob's outcomes, she can estimate the probabilities of his possible outcomes (the second factor in \erf{Supp1}) using her evidence ($A,X,Y$) and the conditional probability
\begin{equation}{\label{BobPr}}
    \text{Pr}(B=b|A=a,X=x,Y=y)=\frac{\wp(a,b|x,y)}{\sum_\beta \wp(a,\beta|x,y)}\,.
\end{equation}
Alice's actual measurement setting and outcomes is taken to be $X=\updownarrow$ and $A=\uparrow$. Since they share a singlet state, this collapses Bob's state to $\ket{\downarrow}$ on which he measures along the rotated basis $Y=$\rotatebox[origin=c]{135}{$\updownarrow$}. 
Using \erf{BobPr}, we have
\beq
\text{Pr}(B=\rotatebox[origin=c]{135}{$\uparrow$}|A=\uparrow,X=\updownarrow,Y= \rotatebox[origin=c]{135}{$\updownarrow$}
)=\frac{1}{2}(1+ \cos{\phi} )=\frac{\sqrt{2}+1}{2\sqrt{2}},
\eeq
where $\phi$ is the angle between the direction of the state and the direction along which it is measured. For Bob's measurement choice, $\phi = \pi/4$. Similarly
\beq
\text{Pr}(B=\rotatebox[origin=c]{135}{$\downarrow$}|A=\uparrow,X=\updownarrow,Y= \rotatebox[origin=c]{135}{$\updownarrow$}
)=\frac{1}{2}(1- \cos{\phi})=\frac{\sqrt{2}-1}{2\sqrt{2}}.
\eeq

In this particular example, Bob's outcomes constitute the classical events that are not in the future light-cone of the event $X'$ (the antecedent), therefore the probabilities of these outcomes serve as probabilities of the unknown fixtures which can now be used in the counterfactual scenario by adding them to the conditional. The probability of her consequent ($A'$) (the first factor in \erf{Supp1}) given her antecedent ($X'$) and for the different possible outcomes of Bob using the conditional probability is thus given by

\begin{equation}{\label{PrAlice}}
    \text{Pr}(A'=a'|B=b,X'=x,Y=y)=\frac{\wp(a',b|x,y)}{\sum_\alpha \wp(\alpha,b|x,y)}\,.
\end{equation} 
%\ktl{I have added a prime on the a, since we use a for the actual value (which is up), and here it is not up.}

Using \erf{PrAlice}, we have
\beq
\text{Pr}(A'=\,\rightarrow|B=\rotatebox[origin=c]{135}{$\uparrow$},X'=\,\leftrightarrow,Y= \rotatebox[origin=c]{135}{$\updownarrow$}
)=\frac{\sqrt{2}+1}{2\sqrt{2}}\,,
\eeq
and
\beq
\text{Pr}(A'=\,\rightarrow|B=\rotatebox[origin=c]{135}{$\downarrow$},X'=\,\leftrightarrow,Y= \rotatebox[origin=c]{135}{$\updownarrow$}
)=\frac{\sqrt{2}-1}{2\sqrt{2}}\,.
\eeq
Summing over all the possible outcomes of Bob, Alice finally computes the probability of her particular consequent $A'=\rightarrow$ in the counterfactual scenario if she had chosen her measurement settings to be $X'=\leftrightarrow$ \textit{given} that she has actually measured in $X=\updownarrow$ and recorded the outcome $A=\uparrow$ to be
\begin{align}
{\rm Su}(A'=\,\rightarrow | X'=\,\leftrightarrow || A=\uparrow,X=\updownarrow,Y=\rotatebox[origin=c]{135}{$\updownarrow$}) \nonumber%\label{CFP}
&\\=\sum_b \Pr(A'=\,\rightarrow |B=b,X'=\,\leftrightarrow,Y=\rotatebox[origin=c]{135}{$\updownarrow$}) \nn 
&\\ \times \Pr(B=b|A=\uparrow,X=\updownarrow,Y=\rotatebox[origin=c]{135}{$\updownarrow$}) \nn
&\\=\frac{\sqrt{2}+1}{2\sqrt{2}}\times\frac{\sqrt{2}+1}{2\sqrt{2}}+\frac{\sqrt{2}-1}{2\sqrt{2}}\times\frac{\sqrt{2}-1}{2\sqrt{2}}=\frac{3}{4}\,. \label{Suppose1}
\end{align}

\subsection{Connecting the CHSH to the general formalism \label{App:A1}} 
See Table~\ref{tab:CHSH_mapping} for the mapping between the CHSH scenario and the general definitions for the counterfactual scenario.

%\newpage
\section{Calculating the average over Bob's possible records \label{App:C}}

In order to compute the final suspectation value in \erf{twotermsinsuspect}, one needs to average over $\both{m}$ with the appropriate probability (second factor), i.e., the conditional probability $\text{Pr}(\both{M}=\both{m}|\both{N}=\both{n})$. This distribution is obtained from the unnormalized state, $\tilde{\rho}^{\past{M},\past{n}}_T$, by using methods developed in quantum state smoothing \cite{PhysRevLett.115.180407,chantasri2021unifying}.
\subsection{Unnormalized states and ostensible distribution \label{App:C1}}

Estimation theory aims to describe the system by conditioning upon measurements. The system takes \textit{different quantum trajectories} or paths, depending on the measurement records $\boldsymbol{R}_t$ obtained in each infinitesimal time $[t+\dd t)$. This process can be described by a set of measurement operations (completely positive maps) $\mathcal{M}_{\boldsymbol{R}_t}$, that evolve the (unnormalized) state forward in time via
\begin{equation}
    \tilde{\rho}_{t+\dd t}^{\past{\boldsymbol{R}}_{t+\dd t}}=\mathcal{M}_{\boldsymbol{R}_t}\tilde{\rho}_t^{\past{\boldsymbol{R}}_{t}}\,,
\end{equation}
 where we introduce the following notation for the records:
\begin{align}
    \past{\boldsymbol{R}}_t=\{\boldsymbol{R}_s:s\in[t_0,t)\}\,, &\\
    \overrightarrow{\boldsymbol{R}}_t=\{\boldsymbol{R}_s:s\in[t,T)\}\,, &\\
    \both{\boldsymbol{R}}_t=\{\boldsymbol{R}_s:s\in[t_0,T)\}\,.
\end{align}
Also note that, when the record has perfect efficiency (no portion of the environment is unobserved), the completely positive map translates to a purity-preserving (mapping pure states to pure states) measurement operation defined as $\mathcal{M}_{\boldsymbol{R}_t} \bullet = \hat{C}_{\boldsymbol{R}_t}\bullet\hat{C}_{\boldsymbol{R}_t}^{\dagger}$. 

In our case there are two observers, Alice and Bob, and we are considering the problem from Alice's perspective. That is, $\boldsymbol{R}=(U,O)$, and we can write $\mathcal{M}_{\boldsymbol{R}_t}\bullet=\mathcal{M}_{U_{t}}\mathcal{M}_{O_{t}}\bullet$ or {\em equiv.}~$\mathcal{M}_{\boldsymbol{R}_t} \bullet =\hat{C}_{\boldsymbol{U}_t}\hat{C}_{\boldsymbol{O}_t}\bullet\hat{C}_{\boldsymbol{O}_t}^{\dagger}\hat{C}_{\boldsymbol{U}_t}^{\dagger}$. Here $U$ is the record unknown to Alice (but observed by Bob), while $O$ is the record observed by Alice.  
The unnormalized state conditioned on both measurement records (unknown and observed) can be calculated by applying a series of measurement operations to the system's initial state $\rho_0$ as
\begin{equation}
    \tilde{\rho}_t^{\past{U},\past{O}}=\mathcal{M}_{U_{t-\dd t}}\mathcal{M}_{O_{t-\dd t}}\ldots\mathcal{M}_{U_{0}}\mathcal{M}_{O_{0}}\rho_0.
\end{equation}
Note that we have omitted the time subscript on $\past{U},\past{O}$ when the context makes it clear.

 Now, given a past observed record $\past{O}$, the correct statistics for the hypothetical past unobserved records $\past{U}$ cannot be obtained by the standard quantum trajectory technique. % (unless we are interested in only a single record of the known and unknown). 
 Since the $\past{U}$ and $\past{O}$ are both `past records' on the total interval $[0,T)$, parts of $\past{O}$ are in the future of parts of $\past{U}$ and so in order to generate $\past{U}$ with the correct statistics given a single $\past{O}$, one needs to consider the later parts of $\past{O}$ that affect the likelihood of the earlier parts of $\past{U}$. This can only be obtained via the unnormalized states, with the unobserved records generated according to an ostensible probability $\wp_{\text{ost}}$, with $U_t$ independent of $U_s$ for $t\neq s$. (See Section~3.3.3 of \crf{chantasri2021unifying} for a more detailed discussion.) Note that this ostensible distribution is something that can be chosen almost arbitrarily, because in the end the evolution of the unnormalized state captures the correct statistics of the unknown records $\past{U}$ obtained from its trace $\text{Tr}[\tilde{\rho}^{\past{U},\past{O}}]$.
 In our case the unobserved and the observed records are Bob and Alice's records, respectively, with
 \begin{align}
     \text{actual}: \both{U},\both{O}=\both{M},\both{N},\nn\\
     \text{counterfactual} :\both{U},\both{O}=\both{M},\both{Y}.
 \end{align}
 
\subsection{The choice of ostensible probabilities for Bob's jump \label{App:C2}} 

Given a record of Alice's photon-detections $\both{N}=\both{n}$ (which in our simulations is just a single jump at time $t_A\equiv 6.25\gamma\inv$), we generate Bob-jumps randomly at a rate $\lambda(t)$. This corresponds to choosing the following ostensible probabilities for the two possible measurement results, i.e., a photon is detected or not, at any  time $t$~\cite{HMWiseman_1996},
\begin{equation}
    \wp_{\text{ost}}(M_t:=1)=\lambda(t) \dd t \;\; \wp_{\text{ost}}(M_t:=0)=1-\lambda(t) \dd t\,.
\end{equation}
It helps to choose the ostensible distribution to be close to the actual distribution in order to maximize the effective ensemble size (and hence minimize the statistical errors)~\cite{chantasri2021unifying}. In this case, the most obvious choice, given Alice's actual record, is the rate that would be generated, at each instant, by Alice's filtered state, namely
\begin{equation}
    \lambda(t)=\gamma\eta_{\rm B}\text{Tr}[\hat{\sigma}_{+}\hat{\sigma}_{-}\rho^{\past{n}}_t], \label{ost}
\end{equation}
and that is what we choose. Note that the technique would not be valid were $\lambda(t)$ to depend on the randomly generated $\past{M}$ in a given run; we cannot replace $\rho^{\past{n}}_t$ in \erf{ost} by $\rho^{\past{M},\past{n}}_t$. 
%which is the ostensible distribution of the unobserved (Bob's records) records and is equal to the rate of the jumps the atom would pertain if the state of the system were Alice's filtered state $\rho^{\past{n}}$. 

% We now show how, using this method of ostensible probabilities, we can generate an ensemble of Bob-records with their actual probabilities, conditioned on Alice's record. \hmw{Is that what you mean? It seems the ensemble is not until quite a bit later, at least as I've moved things around.}
\subsection{Generating an \textit{ostensible}-weighted ensemble of Bob's records given a single record of Alice \label{App:C3}}
Alice samples Bob's records independently in intervals that are determined by her own jump times. This also includes the interval between the initial time $t_0$ and her first recorded jump, and the interval between her final recorded jump and the final time $T$. Restricting Bob’s sampling to these intervals ensures that inconsistencies such as simultaneous jumps recorded by Alice and Bob do not occur. 

With the appropriate choice of the ostensible probabilities, Alice can now generate an ensemble for Bob's photon-detections weighted by this distribution. 
Starting with the initial time $t_0=\tau_0$, Alice can use the cumulative (\textit{ostensible}) probability of a no-jump for Bob to generate a random number $r$ such that it satisfies the equation
\begin{equation}
    r = \exp\ro{-\int_{\tau_0}^\tau \lambda(s) ds}.
\end{equation}
The solution of this equation defines the first time $\tau=s_1$, at which Bob records a jump. Alice repeats the process but this time resetting the initial time $\tau_0=s_1$, to obtain a new solution $s_2$, which corresponds to the second time for Bob to record a jump. This iteration continues, with each step producing an independent jump time within the current interval. In this way Alice generates a random record of jumps for Bob $\{s_1,s_2,\ldots,s_n\}$, in each interval, 
according to the ostensible rate of jumps $\lambda$. 

In the particular case we study, there are only two time intervals: $[t_0,t_A)$ and $[t_A,T)$ determined by her recorded time of jump $t_A = 6.25\gamma\inv$.
%where the number of intervals will be defined by the jumps recorded by Alice.  
By repeating the entire process again, each time from $t_0=0$ to $T=10\gamma\inv$, Alice generates an \textit{ostensible}-weighted ensemble of Bob's records on the interval $[t_0,T)$, which we notate as $ \tilde{\mathfrak{E}} = 
\{\both{M}|\both{N}=\delta_{t,t_A}\}$. Here the tilde emphasises that the elements of the ensemble appear with frequencies given by the ostensible distribution, not the actual distribution. This ensemble comprises two parts: ${ \tilde{\mathfrak{E}}_1} = \{\past{M}_{t_A}|{\both{N}}=\delta_{t,t_A}\}$ and ${\tilde{\mathfrak{E}}_2} = \{ {\fut{M}_{t_A}}|{\both{N}}=\delta_{t,t_A}\}$ with 20,000 Bob's records in each part. Since every record in $\tilde{\mathfrak{E}}_1$ can be combined with every other record in $\tilde{\mathfrak{E}}_2$ the total ensemble size is given by the Cartesian product $|\tilde{\mathfrak{E}}| = |\tilde{\mathfrak{E}}_1| \times |\tilde{\mathfrak{E}}_2|$ which yields an ensemble of size $|\tilde{\mathfrak{E}}|=4\times 10^8$.

\begin{widetext}
\subsection{Obtaining the actual probability of Bob's records from the unnormalized state \label{App:C4}}

The actual probability for each generated record $\both{M}=\both{m}$ in the ensemble $\tilde{\mathfrak{E}}$, is $\Pr[\both{M}=\both{m}|\rho_0,\both{n}] \propto \Tr[\tilde{\rho}^{\past{m},\past{n}}_T]$, where
$\tilde{\rho}_T^{{\past{m}},\past{n}}$ is the unnormalized filtered state at the final time $T$,
conditioned on both Alice's (fixed) and Bob's (varying) past record, corresponding to the above ostensible probabilities for Bob's~\cite{chantasri2021unifying}. The measurement operators for Alice (photon-detections) and Bob (photon-detections) that constitute the map for the evolution of the above mentioned unnormalized filtered state corresponding to the chosen ostensible probabilities are given by
\begin{equation}\hat{C}_{N} \propto\begin{cases}
   1-\frac{1}{2}(\gamma \eta_{\rm A}\hat{\sigma}_+\hat{\sigma}_ -)\dd t   & \text{if no-jump is recorded for Alice }(N=0)\\
\hat{\sigma}_-& \text{if a jump is recorded for Alice }(N=1)
\end{cases},
\end{equation}
and 
\begin{equation}\hat{C}_{M} =\begin{cases}
   1-\left( i\hat{H}+\frac{1}{2}(\gamma\eta_{\rm B} \hat{\sigma}_+\hat{\sigma}_- -\lambda )\right)\dd t   &\text{if Bob records no-jumps } (M=0)\\
\sqrt{ \gamma\eta_{\rm B}/\lambda}\,\hat{\sigma}_-& \text{if Bob records a jump } (M=1)
\end{cases},
\end{equation}
respectively, where $\hat H = (\Omega/2)\s{x}$ is the system Hamiltonian. 

 Evolving under these measurement operators via
 \begin{equation}
   \tilde{\rho}^{\past{M},\past{n}}_{t+\dd t} = \hat{C}_M \hat{C}_{n} \tilde{\rho}^{\past{M},\past{n}}_{t}\blk\hat{C}^\dagger_n \hat{C}_{M}^\dagger\,, 
 \end{equation}
 leads to the (unnormalized) state update
\begin{equation}
    \tilde{\rho}^{\past{M},\past{n}}_{t+\dd t} = 
\begin{cases}
\cu{1 + \dd t\ro{ {\cal L} + \lambda - \gamma\eta_{\rm B}{\cal J}[\hat{\sigma}_{-}]-\gamma\eta_{\rm A}{\cal J}[\hat{\sigma}_{-}]}}\tilde{\rho}^{\past{M},\past{n}}_t  & \text{if no jumps\,,}\\
  (\gamma\eta_{\rm B}/\lambda){\cal J}[\hat{\sigma}_{-}] \tilde{\rho}^{\past{M},\past{n}}_t & \text{if Bob records a jump\,,}\\
\text{arbitrary constant} \times {\cal J}[\hat{\sigma}_{-}] \tilde{\rho}_t^{\past{M},\past{n}}  & \text{when Alice records a jump at} \; (t=t_A)\,,
   
  \end{cases} \label{TS}
\end{equation}
with $\tilde{\rho}_0=\rho_0$, which we choose to be the ground state. Note that averaging~\erf{TS} over Bob's photon-detections yields~\erf{jumptraj}. 

Recall that the records for Bob's jumps in our case were generated independently in the two intervals defined by the jumps of Alice. For a record $\{s_1,s_2,\ldots,s_n\}$ in each interval, \erf{TS} has a piecewise analytical solution given by 
\begin{equation}
    \tilde{\rho}_t^{\overleftarrow{M}, \overleftarrow{n}} =
\begin{cases}
{\rm exp}\left[\int_{t_0}^t \lambda(s) \dd s\right] {\rm exp}\left[{-i \hat{H}_\text{eff}(t-t_0)}\right] \rho_0\, {\rm exp}\left[{i \hat{H}_\text{eff}^\dagger (t-t_0)}\right]  & \text{if } t < s_1, \\
\frac{{\rm exp}\left[{\int_{t_0}^t \lambda(s) \dd s}\right]}{\prod_{k=1}^n \lambda(s_k)} \prod_{i=1}^n A(s_i - s_{i-1}) {\rm exp}\left[{-i \hat{H}_\text{eff}(t-s_n)}\right] \rho_0\, {\rm exp}\left[{i \hat{H}_\text{eff}^\dagger (t-s_n)}\right] & \text{if } t \geq s_1,   
\end{cases} \label{Piecewise}
\end{equation}
where 
\begin{equation}
   A(t) = {\rm exp}\left[{-\frac{\gamma t}{2} } \right]\sin^2 \left( \Omega' t/2 \right) \left( \frac{\Omega}{\Omega'} \right)^2 \eta_{\rm B} \gamma \,\,\,\text{and}\,\,\, \hat{H}_{\text{eff}}(t) = \hat{H} - \frac{i}{2}(\gamma \hat{\sigma}_+\hat{\sigma}_- -\lambda(t)) 
\end{equation} 
and $\Omega' = \sqrt{\Omega^2 - (\gamma/2)^2}$.

The effective Hamiltonian $\hat{H}_{\text{eff}}(t)$ defined above governs the evolution of the atom when there is no jump recorded by either of the observers. Note that it differs from the time-independent $\hat{H}_{\text{eff}}$ by the addition of $i\lambda(t)/2$. 

\end{widetext}
\subsection{Generating an Equiprobable Ensemble of Bob's records by resampling \label{App:C5}} 

The records for Bob, generated according to the ostensible distribution, were sampled independently over the two intervals determined by Alice's recorded jump time $t_A$, $[t_0=0,t_A)$ and $[t_A,T=10\gamma\inv)$. In each interval the unnormalized state $\tilde{\rho}_t^{\overleftarrow{M}, \overleftarrow{n}}$ was evolved until the final time of the interval (using \erf{Piecewise}) for all the 20,000 records in the corresponding part of $\tilde{\mathfrak{E}}$. 
Each record was then assigned a weight equal to trace of the unnormalized state at the end of the interval. It is important to note here that the correct weight in the first interval $[t_0,t_A)$ is, $\text{Tr}\left[\tilde{\rho}_{t_A}^{\overleftarrow{M}, \overleftarrow{n}} \ket{e}\bra{e}\right]$, since Alice records a jump at the end of the first interval. This ensures that Bob's records are conditioned on both the past-future records of Alice,
$\both{n}$. 

The weights were then normalized $(w_1,w_2,\ldots,w_{20,000})$ such that, $\sum_{j=1}^{20{,}000} w_j=1$, ensuring that they can be interpreted as actual probabilities. From these, the marginal sum of the weights  $W_k$, up to the corresponding index $k$, were calculated such that $W_1=0,W_2=w_1,\ldots,W_{20,000}=1-w_{20,000}$. 
A random number $ R \in (0,1]$ was drawn, and the record with largest index $k$ satisfying 
$W_k < R$  was chosen. This procedure was also applied for the second interval in the same run.
In this way a single record for Bob was obtained over the entire time interval $[t_0,T)$ denoted by $\both{M}$. Repeating the process multiple times, a smaller, equal-weighted  ensemble $\mathfrak{E}$, of size $|\mathfrak{E}|=4\times10^4$ was obtained from the \textit{ostensibly}-weighted initial ensemble $\tilde{\mathfrak{E}}$ of size  $|\tilde{\mathfrak{E}}|=4\times10^8$.

%\ktl{That is, via this resampling, we are able to transform $\mathfrak{E} = \{\{\past{M}_{[t_0,6.25)}|\past{N}=\delta_{t,6.25}\}, \{\past{M}_{(6.25,T]}|\past{N}=\delta_{t,6.25}\}\}$ into $\mathfrak{E}' = \{\both{M}|\both{N}=\delta_{t,6.25}\}$.}
%\begin{equation}
%    \tilde{\mathfrak{E}} = \{\past{M}|\past{N}=\delta_{t,6.25}\}\, \rightarrow \,\mathfrak{E} = \{\both{M}|\both{N}=\delta_{t,6.25}\} 
%\end{equation}

\begin{figure}[t]
\centering
\includegraphics[width=0.45\textwidth]{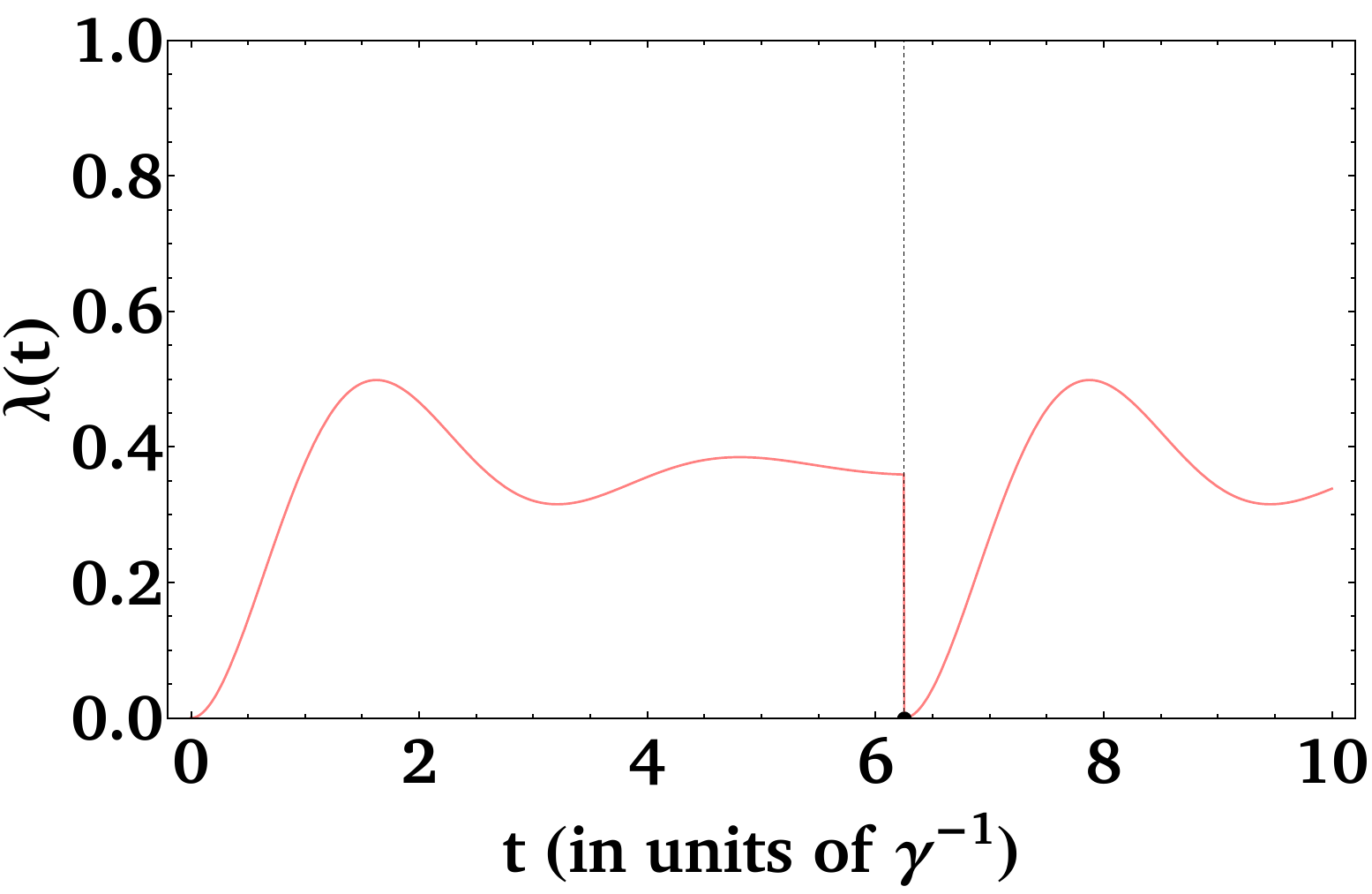}
\hfill
\includegraphics[width=0.435\textwidth]{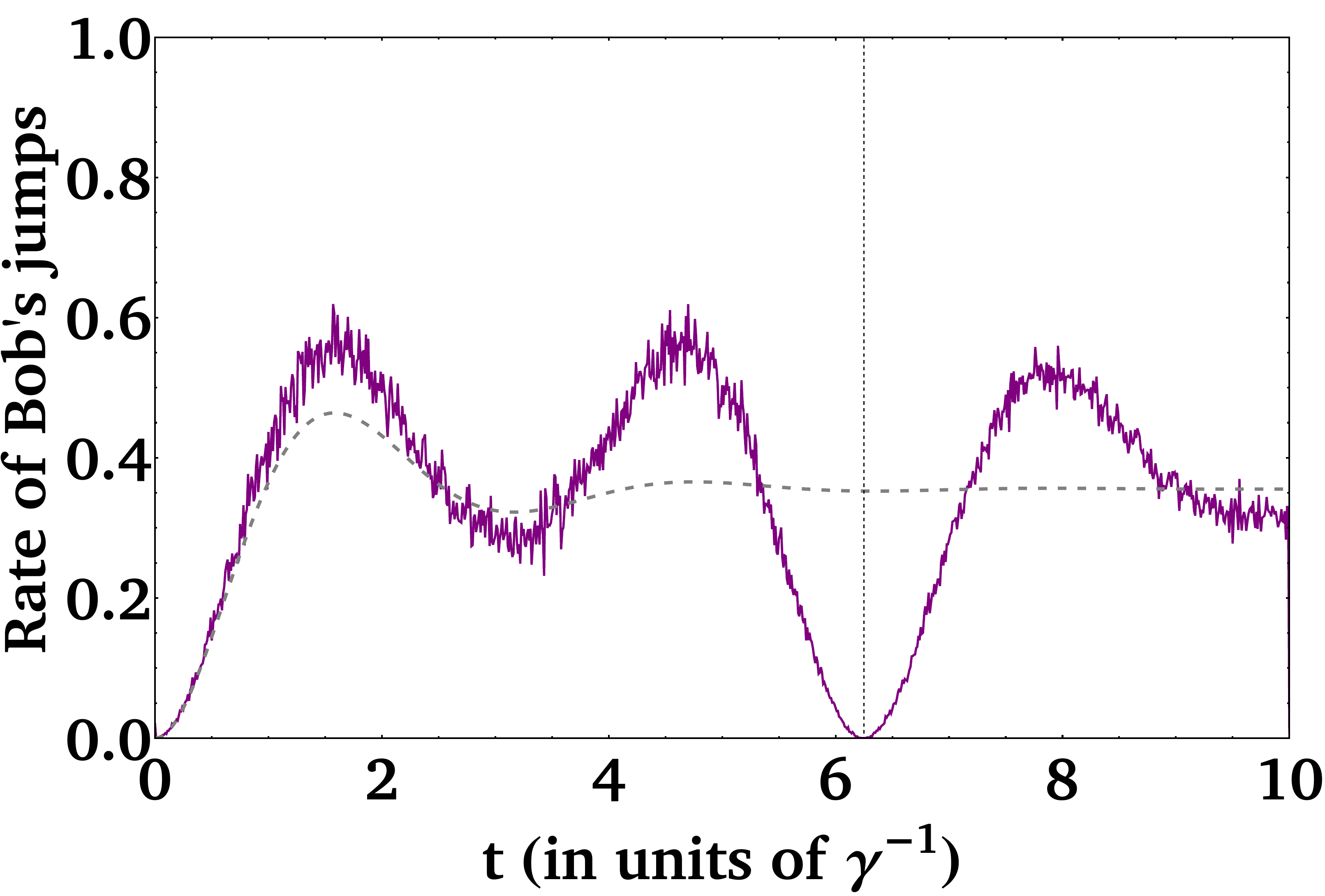}

\caption{
(Top) The ostensible rate of jumps for Bob's detections (pink solid) from \erf{ost} (obtained using Alice's filtered state) which is conditioned only on the past records of Alice ($\past{n}$) and hence drops to zero at $t_A=6.25\gamma\inv$, where Alice detected a photon.
(Bottom) Bob's detection rates (purple solid) in the actual-weighted ensemble $\mathfrak{E}$, that reflects the correct statistics. This is conditioned on Alice's all-time record ($\both{n}$), and due to the additional conditioning on the future records ($\overrightarrow{n}$) it suppresses the rate prior to $t_A=6.25\gamma\inv$ consequently developing a peak at $t=4.71\gamma\inv$. Instead of instantaneously dropping to zero at $t_A$, the distribution now exhibits two peaks roughly half a Rabi cycle before and after this instant. The detection rate from the unconditioned solution (black, dashed) is also plotted for reference.
}
 \label{fig:combined}
\end{figure}

The rate of Bob's detected records in the resampled ensemble $\mathfrak{E}$ is shown in Fig.~\ref{fig:combined} (Bottom). The total time interval is discretized in steps of $\delta t = 0.01$, 
yielding $1001$ time points in total. For each time point, we compute the 
relative frequency of a jump by counting how many records contain that point 
and divide by the total number of records, which is the total size of the ensemble $|\mathfrak{E}|$. The plot of the rate of Bob's jumps reflects how the records were generated with the correct statistics in the resampled ensemble. Specifically, it differs from the ostensible rate $\lambda(t)$ for Bob's jumps, \erf{ost}, which was obtained from Alice's filtered state; this is plotted as a reference in Fig.~\ref{fig:combined} (Top).
When Alice detects a photon at $t=t_A$, her filtered state goes back to the ground state $\ket{g}\bra{g}$ and, correspondingly, the ostensible distribution also goes to zero at this time. This is also seen in the rate of Bob's jumps in $\mathfrak{E}$. 
%\ing{While calculating Bob's rate in FigS1b, we had to take way more records in the resampled ensemble for this convergence, the number 1000 is only for the calculation of the suspectation value.}\hmw{Ah, well you'd better explain that then. And change the sentence after (S.26). You can just introduce the 1000 there.}
 However, unlike $\lambda(t)$, Bob's actual jump rate also is suppressed {\em prior} to the jump. (Thus, in hindsight, we could have chosen a $\lambda(t)$ with this symmetry, and made do with a smaller ensemble $\tilde{\mathfrak{E}}$).  
Going back earlier in time, Bob's jump rate (conditioned on Alice's later detection) has a strong peak at roughly half a Rabi cycle before the minimum. This {\em strong} peak appears neither in $\lambda(t)$ nor in the rate calculated from the unconditioned solution to the master equation $\rho(t)$, also plotted in Fig.~\ref{fig:combined} (Bottom). The point at which the rate of jumps recorded by Bob peaks is at $t=4.71\gamma\inv$, and we use this fact in the next Section to define a ``characteristic'' record for Bob, in order to generate Fig.~\ref{fig:Positive Y}.
% \begin{figure}[H]
%     \centering
%     \begin{subfigure}[b]{0.45\textwidth}
%         \includegraphics[width=\textwidth]{Images/Ostensible distribution.pdf}
%         \caption{}
%         \label{fig:fig1}
%     \end{subfigure}
%     \hfill
%     \begin{subfigure}[b]{0.435\textwidth}
%         \includegraphics[width=\textwidth]{Images/Frequency of Bob's Jumps-2.pdf}
%         \caption{}
%         \label{fig:fig2}
%     \end{subfigure}
%     \caption{(a) (b)}
%     \label{fig:combined}
% \end{figure}

\subsection{Using the Ensemble to calculate the Suspectation value of $Y_t$ \label{App:C6}}

In the counterfactual scenario, Alice's choice of measurement settings is homodyne detection while Bob's setting is fixed as photon-detection. Given we can  generate an equiprobable ensemble of Bob's detections ${\mathfrak{E}}=\{\both{M}|\both{n}\}$, we can replace the probability in any average, like the second factor in~\erf{twotermsinsuspect}, by
\begin{align}
    \sum_{\both m} f(\both{m})\times \blk \text{Pr}(\both{M}=\both{m}|\both{N}=\both{n})  \nn&\\=  \lim_{|{ \mathfrak{E}}|\to\infty} \sum_{\both m \in \mathfrak{E}} f(\both{m})\times \frac{1}{|{ \mathfrak{E}}|}\,,
\end{align}
where $f(\both{m})$ can be any expression depending on $\both{m}$. Thus, given $|{\mathfrak{E}}|\gg 1$, we can approximate the final suspectation value using the equiprobable ensemble as 
\begin{align}
    \mathbb{S}(Y_t |\text{Y-homodyne}\cfg \both{N} = \delta_{t,6.25} )\nn &\\\approx  \frac{1}{|{ \mathfrak{E}}|}\sum_{\both{m} \in\, {\mathfrak{E}}}
 %\ante{\rt{\eta_{\rm A}\gamma}}
 {\text{Re}}\; \frac{\Tr[\hat E^{\fut{m}}_t (i\hat{\sigma}_{-})\hat\rho^{\past{m}}_t]}{{\Tr[\hat E_t ^{\fut{m}}}\hat\rho_t^{\past{m}} ]}\,.
\end{align}
For our calculations of the suspectation value, we do not need such a large ensemble as used to generate \frf{fig:combined} (Bottom). We find that taking $|{\mathfrak{E}}|=1000$ is sufficient to obtain a curve with negligible noise. %  in the counterfactual scenario.
\section{Generating Fig.~\ref{fig:Positive Y}\label{App:D}}

\subsection{General considerations \label{App:D1}}

In this section we detail the steps to generate Fig.~\ref{fig:Positive Y}. That figure was used to help explain the shape of the curve obtained for the suspected value in Fig.~\ref{fig:Results-Hom}. In particular we should explain why, for times $t$ close to $t_A$, the suspected value for Alice's homodyne current in the counterfactual world, conditioned on her detection at $t_A$ in the actual world, is larger than the unconditioned suspectation. [From~\erf{homodynecurrent}, the unconditioned suspected value equals $\Tr[\rho_t\hat{\sigma}_y]$, where $\rho_t$ is the solution to the master equation~\erf{eq:Lindblad}. That is, it is simply an expected value, since a suspectation in a counterfactual world with no evidence from the actual world is simply an expectation in an alternate actual world.] Our explanation for this feature, based on~\erf{changegroundpop}, works if it is the case that $\an{\s{y}}^{\past{Y}}$ is usually positive. But this raises the question of what we mean by $\an{\s{y}}^{\past{Y}}$ and by ``usually''. 

In saying ``usually'' we are recognizing that there is an ensemble of possible values of $\an{\s{y}}^{\past{Y}}$ in the counterfactual world. At the most fine-grained level, we could consider all possible jump records $\both{M}$ for Bob, and, for each of them, all possible Y-homodyne records $\both{Y}$ for Alice. At a more coarse-grained level we could consider only conditioning on $\past{Y}_t$. But these both leave open the question of what distribution of these records to use to define ``usually''. If we were to consider only $\past{Y}_t$, then we could consider $\past{Y}_t$ for any $t$ to be distributed according to the supposability ${\rm Su}(\past{Y}_t||\both{N}=\both{n})$, where $\both{n}$ is the jump at time $t_A$. However, this would seem to be presuming the very thing we are aiming at explaining --- the supposable-statistical properties of $Y_t$ in Fig.~\ref{fig:Results-Hom}. An alternative would be to use an unconditioned distribution for $\past{Y}_t$, but this runs the risk of giving a distribution for $\an{\s{y}}^{\past{Y}}$which is too far removed from what is likely in the counterfactual world to allow valid conclusions to be drawn. 

For these reasons, we here adopt a relatively simple compromise solution in defining the distribution for $\past{Y}_t$. We consider the state conditioned on both $\both{M}$ and $\both{Y}$. But we do not  consider all possible $\both{M}$ according to the distribution 
$\Pr(\both{M}=\both{m}|\both{N}=\both{n})$, and to compute the distribution for $\both{Y}$ in each case, as that would again lead to a distribution for $\both{Y}$ that we are trying to explain. Instead, we pick a single $\both{m}^\chi$ --- a characteristic one given $\both{N}=\both{n}$ --- and from that directly compute the distribution of the filtered-state mean $\Tr[{\rho}_t^{\past{m}^\chi,\past{Y}}\s{y}]$. Note that, unlike in the computation of the suspectation for $Y_t$, at time $t$ this depends only on $\past{m}^\chi_t$, not $\both{m}^\chi$ (still less an ensemble of $\both{m}$).   

In the first subsection below we explain how we generate the distribution (over $\past{Y}_t$) of $\Tr[{\rho}_t^{\past{m},\past{Y}}\s{y}]$, given $\both{m}$. In the second subsection we explain how we choose $\both{m}^\chi$, the most  characteristic $\both{m}$.

\subsection{Generating the ensemble by solving a Fokker--Planck equation \label{App:D2}}
%Now, in the counterfactual scenario we have an equiprobable ensemble of records for Bob ${\mathfrak{E}}=\{\both{M}|\both{n}\}$. 

As just discussed, what we want, given a possible record $\both{m}$ for Bob, is an ensemble of possible Y-homodyne records for Alice such that at any time $t$, $\past{Y}_t$ appears with the correct probability given $\past{m}_t$. 
This ensemble can again be obtained again by using techniques of linear quantum trajectories for an unnormalized state~\cite{chantasri2021unifying}. This unnormalized state, $\tilde{\rho}_t^{\past{m},\past{Y}}$, is conditioned on both the Y-homodyne measurement ($\past{Y}$) and fixed photon-detections for Bob ($\past{m}$). Specifically, it is the solution of a stochastic master equation that, between two consecutive jumps for Bob, has the form
\begin{align}
   \dd\tilde{\rho}_t^{\past{M},\past{Y}} = -i \, \dd t \left[ \hat{H}, \tilde{\rho}_t^{\past{M},\past{Y}} \right] + \gamma \eta_{\rm A}\dd t \,\mathcal{D}[i\hat{\sigma}_-] \tilde{\rho}_t^{\past{M},\past{Y}} \nn\\+ \sqrt{\gamma \eta_{\rm A}}\, \dd W_t \overline{\mathcal{H}}[i\hat{\sigma}_-] \tilde{\rho}_t^{\past{M},\past{Y}} - \gamma\eta_{\rm B}\,\dd t \,   \overline{\mathcal{H}} \left[ \frac{1}{2} \hat{\sigma}_+ \hat{\sigma}_- \right] \tilde{\rho}_t^{\past{M},\past{Y}}, \label{TMSE}
\end{align} 
 where $\overline{\mathcal{H}}[\hat{c}] \bullet = \hat{c} \bullet  + \bullet \hat{c}^\dagger$
represents the linearized version of the operator $\mathcal{H}[\hat{c}] \bullet = \hat{c} \bullet  + \bullet \hat{c}^\dagger  - \text{Tr}(\hat{c} \bullet  + \bullet \hat{c}^\dagger)\bullet$, and $\mathcal{D}[\hat{c}]=\hat{c} \bullet \hat{c}^\dagger-\frac{1}{2}(\hat{c}^{\dagger}\hat{c}\bullet+\bullet\hat{c}^{\dagger}\hat{c})$. Note that averaging~\erf{TMSE} over Bob's photon-detections yields \erf{diffusion}, while averaging it over Alice's homodyne detection leads to the no click evolution of~\erf{S10}.

To be consistent with the actual scenario, the initial state is taken to be the ground state. Since this (pure) unnormalized state is conditioned on all possible records, it will remain pure at all times. In addition to that, the state of the atom is being driven about the $x$-axis,  which  limits it to the $y$--$z$ plane of the Bloch sphere and therefore we can reparametrize it by the Bloch angle $\theta$, such that 
\begin{equation}
    \tilde{\rho}_t^{\past{M},\past{Y}} =(\Lambda_t/2)(\hat{I}+\sin{\theta}\,\hat{\sigma}_y+\cos{\theta}\,\hat{\sigma}_z)\,, \label{ParaTrueState}
\end{equation} 
where, $\Lambda_t\equiv \text{Tr}[\tilde{\rho}_t^{\past{M},\past{Y}}]$ takes care of the state's norm. 

As shown in Section~5.2 of Ref.~\cite{chantasri2021unifying}, one can use this parametrized form of the unnormalized state in \erf{TMSE} to obtain an unnormalized probability density function (PDF) $\tilde{\wp}_{\past{M},\past{Y}}(\theta)$ for $\theta$ given the past records, which gives the normalized PDF simply by 
\begin{equation}
    \wp_{\past{M},\past{Y}}(\theta)=\frac{\tilde{\wp}_{\past{M},\past{Y}}(\theta)}{\int \dd\theta\tilde{\wp}_{\past{M},\past{Y}}(\theta)}. \label{NPDF}
\end{equation}
Specifically, one can obtain $\tilde{\wp}_{\past{M},\past{Y}}(\theta)$ by solving the partial differential equation (PDE)
\beq
\begin{split}
    \partial_t\tilde{\wp}_{\past{M},\past{Y}}(\theta)=&
- \partial_\theta\!\big[ A_\theta(\theta)\, \tilde{\wp}_{\past{M},\past{Y}}(\theta) \big] \\
&- \frac{\gamma \eta_{\rm B}}{2}\,(1 + \cos\theta)\, \tilde{\wp}_{\past{M},\past{Y}}(\theta) \\
&+ \frac{1}{2}\,\partial_\theta^2\!\Big[ B_\theta(\theta)^2\, \tilde{\wp}_{\past{M},\past{Y}}(\theta) \Big]\\
&+ \partial_\theta\!\Big[ B_\theta(\theta)\,\sqrt{\gamma \eta_{\rm A}}\,\sin\theta \, \tilde{\wp}_{\past{M},\past{Y}}(\theta) \Big], \label{PDE}
\end{split}
\eeq
where
\begin{align*}
    A_\theta(\theta)&= 
-\Bigg(
\Omega \;-\; \frac{\gamma  \eta_{\rm B}}{2}\,\sin\theta \;+\; 
\frac{\gamma \eta_{\rm A}}{2}\,\cos\theta \,\sin\theta
\Bigg),\\
\ 
B_\theta(\theta)&= 
- \sqrt{\gamma \eta_{\rm A}}\,\big(1 + \cos\theta\big).
\end{align*}

This can be solved numerically (using Mathematica's `NDSolve' package) with the delta-function initial state approximated by a suitably narrow-width Gaussian function $\wp_{t=0}(\theta)=\frac{1}{\sqrt{2\pi\sigma^{2}}} \exp\,[-\frac{(\theta - \mu)^{2}}{2\sigma^{2}} ]$. For the initial ground state, we have chosen $\theta_0=\pi$ and $\sigma=0.01$, which gives the peak to be $\wp_{t=0}(\pi)\approx39.89$ instead of infinity and appropriate boundary conditions $\wp_{\past{M},\past{Y}}(\theta=0)=\wp_{\past{M},\past{Y}}(\theta=2\pi)\,\forall \,t$.

\subsection{The choice of $\both{m}^\chi$ \label{App:D3}}

%and this  is used in the End Matter to define a ``characteristic'' full record, in which $t=4.71\gamma\inv$ is the time for the last Bob jump prior to the Alice jump at $t_A=6.25\gamma\inv$. This is used as part of the the intuitive explanation for the calculated suspectation value of $Y_t$; see next section.

Recall from Fig.~\ref{fig:combined} (Bottom) that the rate of jumps for Bob, given $\both{N} = \delta_{t,t_A}$, has a maximum of approximately $0.6\gamma$ at $t=4.71\gamma\inv = t_A-1.54\gamma\inv$. The rate then smoothly drops to zero at $t=t_A$. We use these facts to define a 
%To explain the curve of the suspectation value in the main text (Fig.~3) via Fig.~5 in the end matter, we assume a 
characteristic record for Bob ($\past{m}^\chi$) as follows:
\begin{enumerate}
    \item Bob detects a photon at time $t=4.71\gamma^{-1} = t_A-1.54\gamma^{-1}$. We choose this because the rate is maximum here, and he is likely to get a photon within a few $\gamma^{-1}$ of (and prior to) $t_A$, since the maximum rate corresponds to getting a photon roughly every $(0.6\gamma)^{-1} \approx 1.7 \gamma^{-1}$  
    \item Bob does not detect any photons in the interval $(4.71\gamma^{-1},t_A)$.  This is very likely since the raw rate drops monotonically in this interval, as stated, and, more importantly, jumps in resonance fluorescence are anti-bunched.
    \item Other photon-detections can occur at any other time. This is allowed because jumps after $t_A$ have no effect on the filtered state at $t_A$, ${\rho}_{t_A}^{\past{m}^\chi,\past{Y}}$, and jumps before the chosen jump time $t=4.71\gamma^{-1}$ do not alter the conditioned state after that time.
\end{enumerate}  
With the above assumptions we can find the distribution $\tilde{\wp}_{\past{m}^\chi,\past{Y}}(\theta)$ at $t=t_A$ by starting with the initial condition of the distribution $\tilde{\wp}_{\past{m}^\chi,\past{Y}}(\theta)$ being  the narrow-width Gaussian function at $t=t_A-1.54\gamma\inv$. %The time at which Alice detects a jump (the point around which the suspectation value peaks) is $t_A=6.25\gamma\inv$ which is $1.54\gamma\inv$ ahead of $t=4.71\gamma\inv$. 
Hence, we can simply solve the partial differential equation with the above defined $t=0$ state in the preceding subsection up to a time $t=1.54\gamma\inv$. Then we normalize it according to~\erf{NPDF} to obtain the curve for the distribution in Fig.~\ref{fig:Positive Y}.

Finally, in order to obtain the sinusoidal curve of the expectation value of $
\langle\sigma_y\rangle$, one can use the normalized state $\rho_t^{\past{M},\past{Y}}$ (parametrized by the Bloch angle $\theta$ in ~\erf{ParaTrueState}) to obtain
\begin{equation}
    \langle\sigma_y\rangle=\text{Tr}\left[\rho_t^{\past{m}^\chi,\past{Y}}\,\hat{\sigma}_y\right]=\sin{\theta}.
\end{equation}

% \bibliography{References.bib}
%\printbibliography

%

\end{document}